\documentclass[a4paper]{article}
\usepackage{a4wide}
\usepackage{latexsym}   
\usepackage{graphicx,subfigure}
    \graphicspath{{./}{figure/}}
\usepackage{algorithm}
\usepackage{algpseudocode}
\usepackage{epstopdf}
    \graphicspath{{./}{figures/}}
\usepackage[colorlinks=true,linkcolor=blue,citecolor=blue]{hyperref}
\usepackage{xcolor}
\usepackage{amsfonts,amsmath,amsthm,mathtools,bbm,cool}

\newtheorem{theorem}{Theorem}[section]

\theoremstyle{remark}\newtheorem{remark}[theorem]{Remark}

\newcommand{\be}{\begin{equation}}
\newcommand{\ee}{\end{equation}}

\newcommand{\e}{\epsilon}

\newcommand{\fer}[1]{(\ref{#1})}
\newcommand{\R}{\mathbb R}
\newcommand{\N}{\mathbb N}
\allowdisplaybreaks
\begin{document}
\title{Control of tumor growth distributions\\ through kinetic methods}

\author{Luigi Preziosi 
        \thanks{Department of Mathematical Science ``G. L. Lagrange'', Politecnico di Torino, Italy. {\tt luigi.preziosi@polito.it}} \and
        Giuseppe Toscani \thanks{Department of Mathematics ``F. Casorati'', University of Pavia, and The Institute for Applied Mathematics and Information Technologies of CNR, Pavia, Italy. {\tt giuseppe.toscani@unipv.it}} \and
         Mattia Zanella \thanks{Department of Mathematics ``F. Casorati'', University of Pavia, Italy. {\tt mattia.zanella@unipv.it}} 
        }  
\date{}

\maketitle

\begin{abstract}
{The mathematical modeling of tumor growth has a long history, and has been mathematically formulated in several different ways. Here we tackle the problem in the case of a continuous distribution using  mathematical tools from  statistical physics. To this extent, we introduce a novel kinetic model of growth which highlights the role of microscopic transitions in determining a variety of equilibrium distributions. At variance with other approaches, the mesoscopic description in terms of elementary interactions allows to design precise microscopic feedback control therapies,  able to influence the natural tumor growth and to mitigate the risk factors involved in big sized tumors.  We further show that under a suitable scaling both the free and controlled growth models correspond to Fokker--Planck type equations for the growth distribution with  variable coefficients of diffusion and drift, whose steady solutions in the free case are given by a class of generalized Gamma densities which can be characterized by fat tails. In this scaling the feedback control produces an explicit modification of the drift operator, which is shown to strongly modify  the emerging distribution for the tumor size. In particular, the size distributions in presence of therapies manifest slim tails in all growth models,  which corresponds  to a marked mitigation of the  risk factors. Numerical results confirming the theoretical analysis are also presented. }
\medskip

\noindent{\bf Keywords:} kinetic modelling; tumor growth; control \\

\noindent{\bf Mathematics Subject Classification:} 35Q20; 35Q92; 35Q93
\end{abstract}

\tableofcontents
\section{Introduction}\label{sec:intro}
Since the early years of cancer research one of the basic questions addressed by
scientists aimed at the identification of the growth law followed by tumors. The natural related purpose was the need of using it to model the effect of cancer treatment and optimize therapy.

{The easiest but still most used way to do that is to model growth by an ODE, usually of first order. }According to the right hand side, they are named after Malthus (i.e., the exponential growth law), Verhulst (i.e., logistic growth law), Gompertz, Richards, von Bertalanffy, West, and so on.
{In particular, West et al. \cite{West} gave a new insight to von Bertalanffy's growth model starting from an original viewpoint. The parameters of the model are then optimized to fit the available experimental data in absence and presence of therapies. Due to the need of fitting the same data, they mostly give rise to a similar sigmoidal behaviour characterized by an asymptotic tendency to an equilibrium related to the presence of a carrying capacity. } The literature on the subject is huge. So, for more information we refer to the recent review papers \cite{Gerlee, Rodriguez, Sarapata} and volumes \cite{Ledzewich, Wodarz}.

{The process of parameter identification is affected by many sources of uncertainty stemming out at different and independent levels of observation.} To name a few, the first one consists in the fact that the evaluation of the number of cells in a tumor is obtained using only partial information, e.g., approximating the tumor as an ellipsoid on the basis of the maximum and the minimum dimension measured ex-vivo (the middle axis of the ellipsoid is then approximated as the mean of the measurements above), or obtained by two-dimensional in-vivo images assuming that the observed section is the one containing the longest and shortest axis of the ellipsoid. The second one regards the presence within the same body of many metastasis of different sizes growing in different environmental conditions. The third regards the fact that in a cohort of individuals, from nude mice used in experiments up to humans, the evolution is not the same because in each host the response of the body is different. 

%{ 
%Exploiting similarities between phenomena belonging to different fields is a powerful strategy for the mathematical modeling of new phenomena, and to give a new insight into them. As far as the cancer growth models are concerned (cf. the recent review \cite{Gerlee}), a new insight into the von Bertalanffy's growth has been recently given by West et al., who proposed a general model for ontogenetic growth \cite{West}}

So, in spite of the apparent simplicity of the question, at present there is no general consensus on the type of growth law that is better to be used to fit data, with stochasticity playing a role that is often overwhelming with respect to the difference among the evolutions predicted by the different models. {In addition, the relation between the therapeutic action operating at the cellular level and the macroscopic parameter, e.g. the carrying capacity, is not immediate.} On the other hand, regardless of the exact fitting of the growth law, as stated for instance in \cite{Folkman}, one of the therapeutic goals in oncology is to control tumor growth and to reduce the probabilities of having tumors growing to sizes that are too large to be physiologically or therapeutically controllable, or that are harmful to the human body. 

In order to accomplish this task, rather than modelling the tumor with a stochastic adaptation of the ODE growth models, we present here a {novel kinetic approach, which aims to describe the growth of tumor cells in terms of the evolution of a distribution function whose temporal variation is the result of transitions occurring at the cellular level that lead to an increase or decrease in tumor size, related to growth and death processes. The mathematical description proposed here is based on a Boltzmann-type model where the elementary variations describing the number of cancer cells are determined by a transition function which takes environmental cues and random fluctuations into account. Under a suitable limit procedure, different choices of parameters in the transition probability will characterize the equilibrium distribution. }

The notion of growth in random environment has been formulated before in the framework of stochastic birth and death processes by several Authors (see, for instance, \cite{Nobile, Prajneshu, Tan} and references therein) to take into account of environmental fluctuations. In this framework, a  stochastic model of tumor growth was introduced by \cite{AG}.

{Application of tools from nonlinear statistical physics to describe biological phenomena involving a huge number of entities represents one of the major challenges in contemporary mathematical modeling \cite{Adam,Bellomo,BD,BLM,Perth}. A consistent part of these applications makes a substantial use of  methods borrowed from kinetic theory of rarefied gases, which, starting from a mesoscopic description of microscopic cells interactions, leads to construct master equations of Boltzmann type, usually referred to as kinetic equations, able to drive the system towards  universal statistical profiles. An important example of emergent behavior is concerned with the building of tumors by cancer cells and their migration through the tissues \cite{BD,GC,HBS,MD}.  Another  example to consider in this context is the classical Luria--Delbr\"uck mutation problem treated by statistical methods \cite{KP, Tos4}. 
In multicellular organisms, these two examples are closely related, since the connection between mutagenesis and carcinogenesis is broadly accepted (see, for instance, \cite{Frank1,Frank2,Ken}). Furthermore, the Luria--Delbr\"uck distribution plays an important role in the study of cancer, because tumor progression depends on how heritable changes (mutations) accumulate in cell lineages.  While the basic entities in these examples differ from the physical particles in that they already have an intermediate complexity,  for some specific phenomena, like the statistical growth of mutated cells, one can reasonably assume that the statistical behavior of the system is mainly related to the peculiar way entities interact and not to their internal complex structure. 

%A non secondary advantage of the mesoscopic modeling in terms of elementary interactions, is the possibility to take advantage of the similarities that the underlying elementary interactions exhibit in correspondence to phenomena which occur in apparently disconnected fields. Indeed, the microscopic variations we will consider in this paper are coherent with other known growth models in suitable ranges of parameters. Recent advances in this direction are related to the formation of lognormal distribution in human behaviors \cite{DT,DT2,GT1,GT2} and inspired by early socio-economic considerations \cite{KT}.  
}

{ From our point of view, the most important output of using a kinetic model is to have an equilibrium distribution stemming from stochastic interactions occurring at the microscopic level, i.e. the cellular level. In particular, the distribution function will give the probability of having tumours of size bigger than a given alerting size. Most importantly, it will be shown that}, in different regimes of parameters of the general transition law, the emerging equilibrium distribution of the { Boltzmann-type} model shows a radically heterogeneous behavior in terms of the decay of the tails. In details, {transitions laws that in a suitable limit are related to} logistic-type growths are associated to a generalized Gamma density function which is characterized by slim tail, i.e. by exponential decay. On the other hand, {transitions laws that in a suitable limit are related to } von Bertalanffy-type growths are associated to Amoroso-type distributions that are rather characterized by fat tail, i.e. by polynomial decay. The border case between the two distributions leads to lognormal-type equilibria which exhibits  slim tail, but with a possible dramatic increase of higher moments.  From a statistical physics point of view, it is worth to remark that in the context of tumor growth the dynamics leading to fat-tailed distributions imply the formation of big sized tumors with high probability. Therefore, the distributions with fat tails can be associated to an increased risk for the human body.
 
{For this reason, once characterized the emerging distributions of the mentioned growth dynamics, we concentrate on implementable therapeutical control strategies so that fat tails can be transformed in thin tails which means mitigating the risk of having big tumors. The control is determined analytically for any growth law. }The control of emerging phenomena described by kinetic models or mean field theories is relatively recent \cite{AHP,APTZ,APZ14,BFY,DHL,FPR}. In particular, the proposed approach can be derived from a model predictive control (MPC)  strategy which is based on determining the control by optimising a given cost functional over a finite time horizon which recedes as time evolves \cite{CB,Son}. Assuming that the the minimisation horizon coincides with the duration of a single transition, we obtain a feedback solution to the control problem that can be implemented efficiently in the Boltzmann-type kinetic model to observe its aggregate effects. It is well known that MPC leads typically to suboptimal controls. Nevertheless, performance bounds are computable to guarantee the consistency of the MPC approximation in a kinetic framework \cite{Grune,HZ}. 

%We will prove that the ultimate effect of therapeutical protocols, which we mimic through control methods, relies on a strong modification of the emerging distribution for the tumor size. In particular, the size distributions in presence of therapies manifest slim tails in all growth models, which should mitigate the aforesaid risk factors. 

In more detail, the paper is organized as follows. 
%in Section \ref{sec:micro} we {  briefly review the standard approach (cf. \cite{AG} for further details), starting from the description of the main}  microscopic growth models and  {  their coupling with random fluctuations, which lead to Fokker-Planck type equations which describe} the dynamics of the statistical growth. 
The kinetic model for tumor growth is presented in Section \ref{kinetic} where we introduce elementary variations of the number of cancer cells depending on a transition function determining the deterministic variations of the tumors' size, and on random fluctuations. In suitable regimes we will obtain a classification of equilibrium distributions corresponding to the introduced growth models, some of them exhibiting fat tails. The controlled model is presented in Section \ref{control} and the emerging slim tailed distributions are computed for two possible therapeutical strategies. Finally, we summarise the highlights of the work and draw some conclusions. {In Appendix \ref{app:A}-\ref{app:B}  we present a brief review of microscopic and mean-field models for tumour growth.}

\section{Kinetic modeling of tumor growth}\label{kinetic}

\subsection{The kinetic description}\label{grazing}

  {As recalled in Section \ref{sec:intro},  tools from  statistical physics are widely used to to describe biological phenomena involving a huge number of entities. In more details, we} aim to model the statistical growth of metastatic tumors in a population of patients or animals by means of the approach of kinetic theory of multi-agent systems \cite{PT2}.   {The leading idea of kinetic theory is to express the dynamics of the distribution of a certain phenomenon in terms of the microscopic process ruling its elementary changes. In the case under investigation, the phenomenon to be studied is the growth process of cancer cells, which we assume to be measured by a variable $x$ representing the number of diseased cells which varies with continuity in $\mathbb R_+$.  }
  {
In other words, if $X(t)$ denotes the random variable expressing the number of cancer cells at time $t \ge 0$, subject to the initial condition $X(0)=1$, $f(x,t)$ is the probability density associated to the process $X(t)$ such that $f(x,t)dx$ is the fraction of tumours which, at time $t\ge 0$ are characterized by size  between $x$ and $x + dx$.  In recent years mean-field approaches have been developed in the field, we summarise the main ideas in Appendix \ref{app:B}. }
 
  {
Following the well-consolidated approach developed in last decade \cite{FPTT, NPT, PT2} in the context of interacting systems, the study of the time evolution of the probability density $f(x,t)$ of cancer cells, together with a reasonable explanation of the growth process induced by this distribution, can be achieved by means of kinetic models. }Then, the knowledge of $f(x,t)$ allows to compute the evolution of aggregate quantities of interest. In particular, since
\[
\int_{\R_+} f(x, t)\, dx = 1,
\]
 for any given smooth function $\varphi(x)$ (the observable), the quantity
\[
\int_{\R_+} \varphi(x) f(x, t)\, dx.
\]
provides the evolution of an observable quantity. Important observable quantities are the principal moments of the density, $\varphi(x) = x^n$, $n \ge 1$, and the characteristic function $\chi_A(x)$ of an interval $A\subseteq \R_+$, whose evolution quantifies the percentage of cancers with a number of cells, $x \in A$ at time $t \ge 0$.   {For instance, if $\varphi(x) = x$ one has the evolution of the mean size of tumours. }

In agreement with the classical kinetic theory of rarefied gases, which aims at describing the dynamics of a huge number of particles, we assume that   {the evolution in time of the density $f$ is due to repeated microscopic interactions which modify the size $x$ of the tumor}. In the sequel, let $x_L$ denote the mean number of cells that can be reached with nutrients in the type of tumor under consideration. 

 For any given value $x$ of cancer cells, we model the elementary variation $x \to x^\prime$  as follows
\begin{equation}\label{eq:inter1}
x^\prime = x + \Phi^\epsilon (x/x_L) x + x\eta_\e, 
\end{equation}
{ where $x_L$ is the characteristic tumour size, e.g. the carrying capacity. }
Thus, in a single transition the tumor's size $x$  can be modified by two different mechanisms, expressed in mathematical terms by two multiplicative terms, both parameterized by a small positive parameter $\e \ll 1$, quantifying the intensity of the interaction itself:
\begin{itemize}
\item[$i)$] the transition function $\Phi^\epsilon (\cdot)$ characterizes the small deterministic variations of the tumors' size, as a function of the quotient $x/x_L$, due to environmental cues. 

\item[$ii)$] the random variable $\eta_\epsilon$ characterizes the  fluctuations due to unknown factors.  The usual choice is to consider that the random variable $\eta_\e$ is of zero mean and variance   {of the order of $\epsilon$}, expressed by $\langle \eta_\e \rangle =0$, $\langle \eta_\e^2 \rangle  = \e\sigma^2$. 
\end{itemize}
  {Starting from the elementary interaction \fer{eq:inter1}, and resorting to classical kinetic theory  \cite{PT2},  the time-evolution of the statistical distribution $f(x,t)$ of the number of cancer cells can be described by a kinetic master equation}.  Indeed, the elementary transition process \fer{eq:inter1} induces a time variation of the  density $f(x,t)$  which is quantified by a linear Boltzmann-type operator. The corresponding kinetic equation is fruitfully written in weak form  \cite{Cer, PT2}. The weak form corresponds to say that the solution $f(x,t)$
satisfies, for all smooth functions $\varphi(x)$ determining observable quantities
\begin{equation}\label{eq:boltz_1}
\dfrac{d}{dt} \int_{\mathbb R_+} \varphi(x) f(x,t)dx = \left \langle \int_{\mathbb R_+}   {B(x)} (\varphi(x^\prime)-\varphi(x))f(x,t)dx \right\rangle,
\end{equation}
where with $\langle \cdot \rangle$ we denoted the expectation with respect to the random parameter $\eta_\e$ introduced in \fer{eq:inter1}.   {In \fer{eq:boltz_1} the function $B(\cdot)$ is a kernel characterizing the  frequency of the elementary growth transitions in presence of tumour cells of size $x$. }
%In the following we will always assume $B \equiv 1$, meaning that the frequencies of growth transitions do not depend on the actual number of tumour cells, leaving to future works the detailed study of effects of nonconstant kernels on the evolution of the distribution $f$.

The right-hand side of equation \fer{eq:boltz_1} represents the variation of the mean value of the observable quantity $\varphi(\cdot)$  consequent to growth transitions that, according to \fer{eq:inter1}, modify  the number of cancer cells from from the { pre-transition} value $x$ to the { post-transition} value $x^\prime$. 

  {Letting $\varphi(x) = 1$ in \eqref{eq:boltz_1} we obtain
\[
\dfrac{d}{dt} \int_{\mathbb R_+} f(x,t)\, dx = 0,
\]
hence, the introduced kinetic model, for any $B(\cdot)$,  is such that
 \[
\int_{\mathbb R_+} f(x,t)\,dx = \int_{\mathbb R_+} f(x,t=0)\,dx.
 \]
 In reason of this we notice that the mapping $x \mapsto f(x,t)$ as a probability density function on $\mathbb R_+$ for all time $t \ge 0$ is compatible with this property of equation \eqref{eq:boltz_1}. At difference with this simple case,
the precise computations of the evolution of higher moments, which correspond to the choice $\varphi(x) = x^n$, with $1 \le n \in \N$, appears cumbersome, and in any case impossible to express analytically. For this reason, it is fruitful to introduce into the integral transition operator in \fer{eq:boltz_1}  some simplifications, which consist first in choosing a unitary kernel, $B(x)\equiv  \textrm{constant}$, and second in considering a suitable scaling for the model parameters.}

  {
It is worth to remark that the choice of a frequency kernel $B(\cdot)$ independent of the number of tumor cells, meaning that the frequencies of growth transitions do not depend on the actual number of tumour cells, does not modify the features of the equilibrium configuration \cite{FPTT1}. Therefore, even if this choice may appear simplistic for modelling purposes, it does not influence the forthcoming analysis. We leave to future works the detailed study of effects of variable kernels on the dynamics of the distribution $f$. }
%  Also, the scaling of the kinetic master equation leads to a Fokker--Planck type model where the probability density changes accordingly to a continuous set of states, and where various results can be treated analytically.  

  {Furthermore, we will adopt the so-called \emph{quasi-invariant} scaling for the introduced transition function,  meaning that the transitions of the proposed model can be considered arbitrarily small. This is expressed by  assuming that $\Phi^\epsilon (\cdot)$ is of the order of $\epsilon$, and that
\be\label{scala}
\lim_{\epsilon \to 0^+} \frac{\Phi^\epsilon (x/x_L)}{\epsilon}  = \Phi(x/x_L).
\ee
The main idea behind the quasi-invariant scaling is to fix, for a given choice of $\e \ll 1$  in \eqref{eq:inter1} a $\e$-dependent value of the frequency $B$ balancing the smallness of the single transition by increasing its frequency and to obtain, in correspondence to any observable quantity $\varphi(\cdot)$, a non vanishing variation of its mean value even in the limit $ \e\to 0 $.  As shown in \cite{FPTT}, where the computations are presented in full details, the right correction for the kernel is to multiply it by $1/\e$.  An analogous effect is obtained by introducing the time scale $\tau = \e t$ such that in correspondence of $\e \rightarrow 0^+$ we consider the large time behavior of the system at time $t \rightarrow +\infty$, meaning that since the contribution of the single transition is small, we need to wait enough time to observe changes as $\e \to 0^+$. It is worth to mention that the quasi-invariant scaling overrides the role of \emph{grazing} interactions \cite{PT2,Tos,Vil}. }  

Hence, let us fix into \fer{eq:boltz_1} the kernel $B = 1/\e$.  Then, $f$ satisfies the equation
\begin{equation}\label{eq:boltz2}
\dfrac{d}{d t}  \int_{\mathbb R_+} \varphi(x)  f(x,{t})dx = \dfrac{1}{\epsilon}\left \langle \int_{\mathbb R_+} (\varphi(x^\prime)-\varphi(x))  f(x, t)dx \right\rangle.
\end{equation}
Since if $\epsilon \ll 1$, the difference $x^\prime -x$ is small, and assuming $\varphi$ sufficiently smooth and rapidly decaying at infinity, we can perform the following Taylor expansion
%{\bf (Controlla qui che ho fatto delle correzioni e non so se sono giuste)}
\[
\varphi(x^\prime) -\varphi(x) = (x^\prime - x)\partial_x \varphi(x) + 
\dfrac{1}{2} (x^\prime - x)^2\partial_x^2\varphi(x) + \dfrac{1}{6}(x^\prime-x)^3 \partial_x^3 \varphi(\bar x), 
\]
being $\bar x \in \left( \min\{x,x^\prime\},\max\{x,x^\prime\} \right)$. Writing $x^\prime - x = \Phi^\e(x/x_L)x + x\eta_\e$ from \eqref{eq:inter1} and plugging the above expansion in \eqref{eq:boltz2} we have
\begin{equation}
\label{eq:4star}
\begin{split}
&\dfrac{d}{d{t}} \int_{\mathbb R_+} \varphi(x) f(x,{t})dx  \\
&\qquad= \dfrac{1}{\epsilon} \left[\int_{\mathbb R_+}  \Phi^\e(x/x_L)x \partial_x \varphi(x) f(x,{t}) dx + \dfrac{\sigma^2   {\epsilon}}{2} \int_{\mathbb R_+}\partial_x^2 \varphi(x) x^2 f(x,{t})dx \right] + R_\varphi(f)(x,{t}),
\end{split}
\end{equation}
where $R_\varphi(f)$ is the remainder
\[
\begin{split}
R_\varphi(f)(x,{t}) &=  \dfrac{1}{2\epsilon} \int_{\mathbb R_+}\partial_x^2 \varphi(x)\left(\Phi^\e(x/x_L)\,x\right)^2 f(x,{t})\, dx \\
&+ \dfrac{1}{6\epsilon} \left\langle \int_{\mathbb R_+} \partial_x^3\varphi(\bar x) \left( \Phi^\e(x/x_L)x + x\eta_\e \right)^3 f(x,{t})dx \right\rangle. 
\end{split}\]
By assumption, $\varphi$ and its derivatives are bounded in $\mathbb R_+$ and rapidly decaying at infinity. Further, if $\eta_\e$ has bounded moment of order three, namely $\langle |\eta|^3 \rangle <+\infty$, using the bound \fer{scala}  we can easily argue that in the limit $\epsilon \rightarrow 0^+$ we have
\[
\left| R_\varphi(f) \right| \rightarrow 0, 
\]
Hence, in the limit $\epsilon \rightarrow 0^+$ equation \eqref{eq:4star} converges to
\[
\dfrac{d}{d{t}} \int_{\mathbb R_+} \varphi(x) f(x,{t})dx = \int_{\mathbb R_+}  { \Phi\left(\frac x{ x_L}\right)} x  f(x,{t}) \partial_x\varphi(x) dx + \dfrac{\sigma^2}{2} \int_{\mathbb R_+} x^2  f(x,{t}) \partial_x^2\varphi(x)dx. 
\]
  {If for any given $t \ge 0$ the limit density $f=f(x,t)$ satisfies, at the point $x =0$, the no-flux boundary condition
\[
\begin{split}
  {- \Phi\left(\frac x{ x_L}\right)} x f(x,{t}) +\dfrac{\sigma^2}{2} \partial_x (x^2 f(x,{t})) \Bigg|_{x = 0} =0,
%x^2 f(x,{t}) \Bigg|_{x = 0} = 0,
\end{split}
\]
integrating  by parts we conclude that  it solves the kinetic equation
\begin{equation}\label{eq:weakFP1}
\dfrac{d}{dt} \int_{\mathbb R_+} \varphi(x) f(x,t)\, dx =  \int_{\mathbb R_+} \varphi(x)  \partial_x \left[  { -  \Phi\left(\frac x{ x_L}\right)} x f(x,{t}) +\dfrac{\sigma^2}{2} \partial_x (x^2 f(x,{t}))\right]\, dx,
\end{equation}
that corresponds to the weak form of } a Fokker-Planck equation (in divergence form)
\begin{equation}\label{eq:FP}
\partial_{t} f(x,t) = \partial_x \left[  { -\Phi\left(\frac x{ x_L}\right)} x f(x,{t}) +\dfrac{\sigma^2}{2} \partial_x (x^2 f(x,{t}))\right],
\end{equation}
with variable coefficient of diffusion, and drift term characterized by the limit function $\Phi(\cdot)$ defined in \fer{scala}.
  {Both the Boltzmann-type equation \fer{eq:boltz_1}  and the Fokker--Planck type equation \fer{eq:FP} describe the evolution in time of the statistical distribution of tumor growth consequent to the microscopic interaction \fer{eq:inter1}. These two kinetic equations, which correspond to different intensities of interactions, allow to obtain an exhaustive description of both the evolution and the stationary statistical states. In particular, the Boltzmann type description will be at the basis of the study of optimal control strategies of the statistical growth.}

\subsection{Transition functions and elementary growth}

  {In this Section, we will detail the elementary interaction \fer{eq:inter1} by introducing  a class of transition functions $\Phi^\epsilon(\cdot)$, which can be easily linked to different growth models. We point the interested reader to Appendix \ref{app:A} for an insight on microscopic modeling of tumor growth, and for a better understanding of the forthcoming analysis.} 
{The function $\Phi^\epsilon (\cdot)$ will be chosen in a way  to properly characterize  the elementary mechanism of growth of the tumor under consideration. Clearly, the function $\Phi^\epsilon (\cdot)$ has to be positive when $x <x_L$, thus producing, in absence of random fluctuations, a growth of the value $x$, and decreasing on this interval, in agreement with the fact that tumour growth slows down for values of $x$ in proximity of the carrying capacity. Moreover, if to avoid unnecessary mathematical \emph{cut-off} assumptions, we extend the interval of possible values assumed by the function $\Phi^\epsilon (\cdot)$ to the whole of $\R_+$, the function on the interval $x>x_L$ has to be negative and slowly increasing,  thus expressing an almost negligible possibility for the number $x$ of tumor cells  to cross the carrying capacity value $x_L$. 

A  transition function with these characteristics is given by 
\be\label{Phi0}
 \Phi_0^\epsilon(s) = \mu\, \frac{1-s^\e }{(1+\lambda) s^\e + 1-\lambda}, \quad s \in \R_+,
 \ee
where  $0 < \e \ll 1$, $0 < \mu <1$ and $0 \le \lambda<1$ are given non negative constants. In agreement with the previous observations, for any given value of these constants, the function $\Phi_0^\epsilon(s)$ is decreasing and convex on $\R_+$, and equal to zero at the reference point $s=1$, where $x =x_L$. Moreover it takes values in the interval
 \be\label{b+-}
 -\frac\mu{1+\lambda}  \le \Phi_\delta^\e(s) \le \frac\mu{1-\lambda},
  \ee
that does not depend on the parameter $\e$. Consequently,   the values of $\mu$ and $\lambda$ characterize the maximal amount of birth and death of tumor cells in a single interaction. Note that the lower bound in \fer{b+-} guarantees that  the deterministic part of the post-interaction value remains positive, since $\mu/(1+\lambda) < 1$. 
\begin{figure}
\centering
\includegraphics[scale = 0.35]{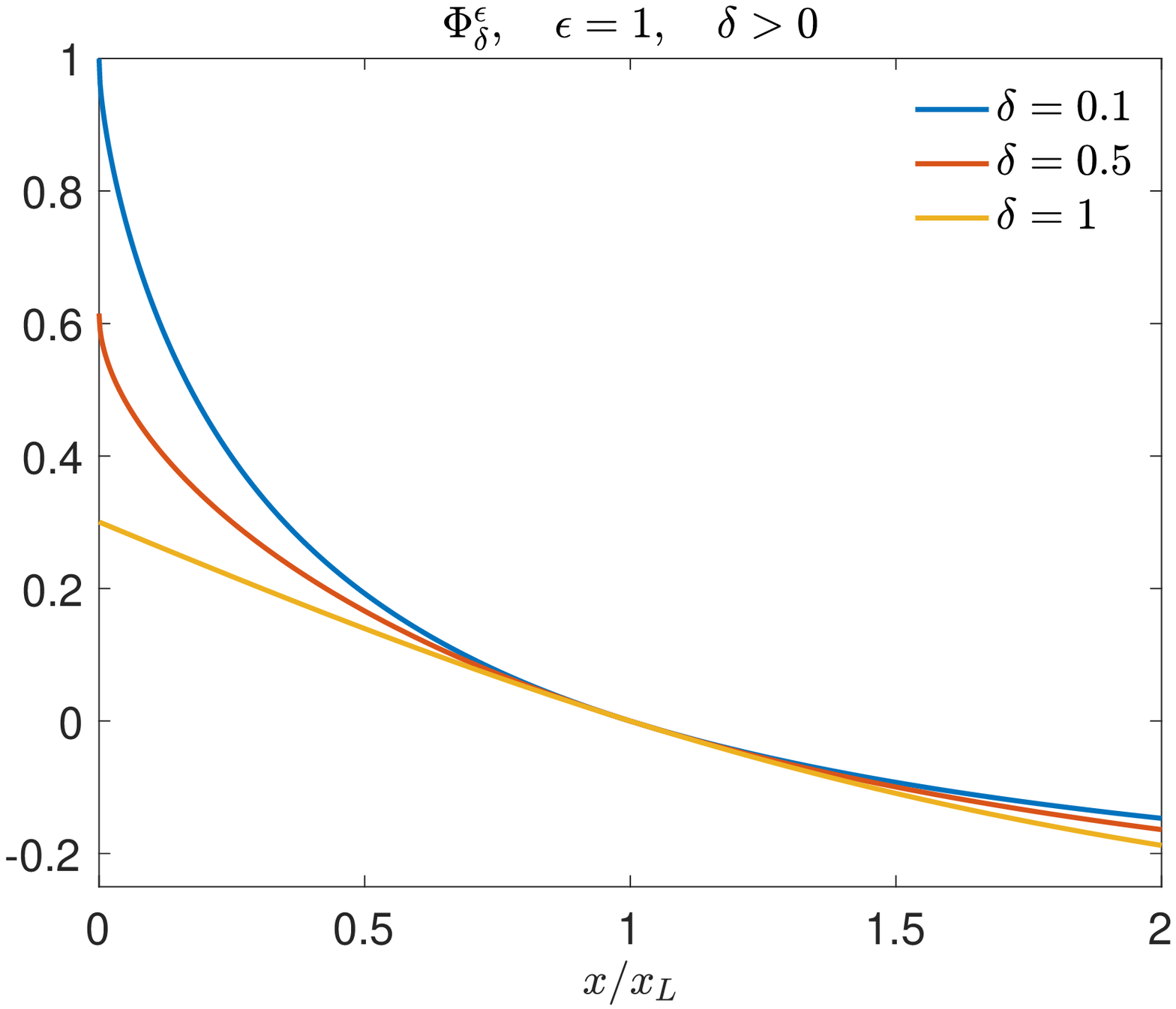}
\includegraphics[scale = 0.35]{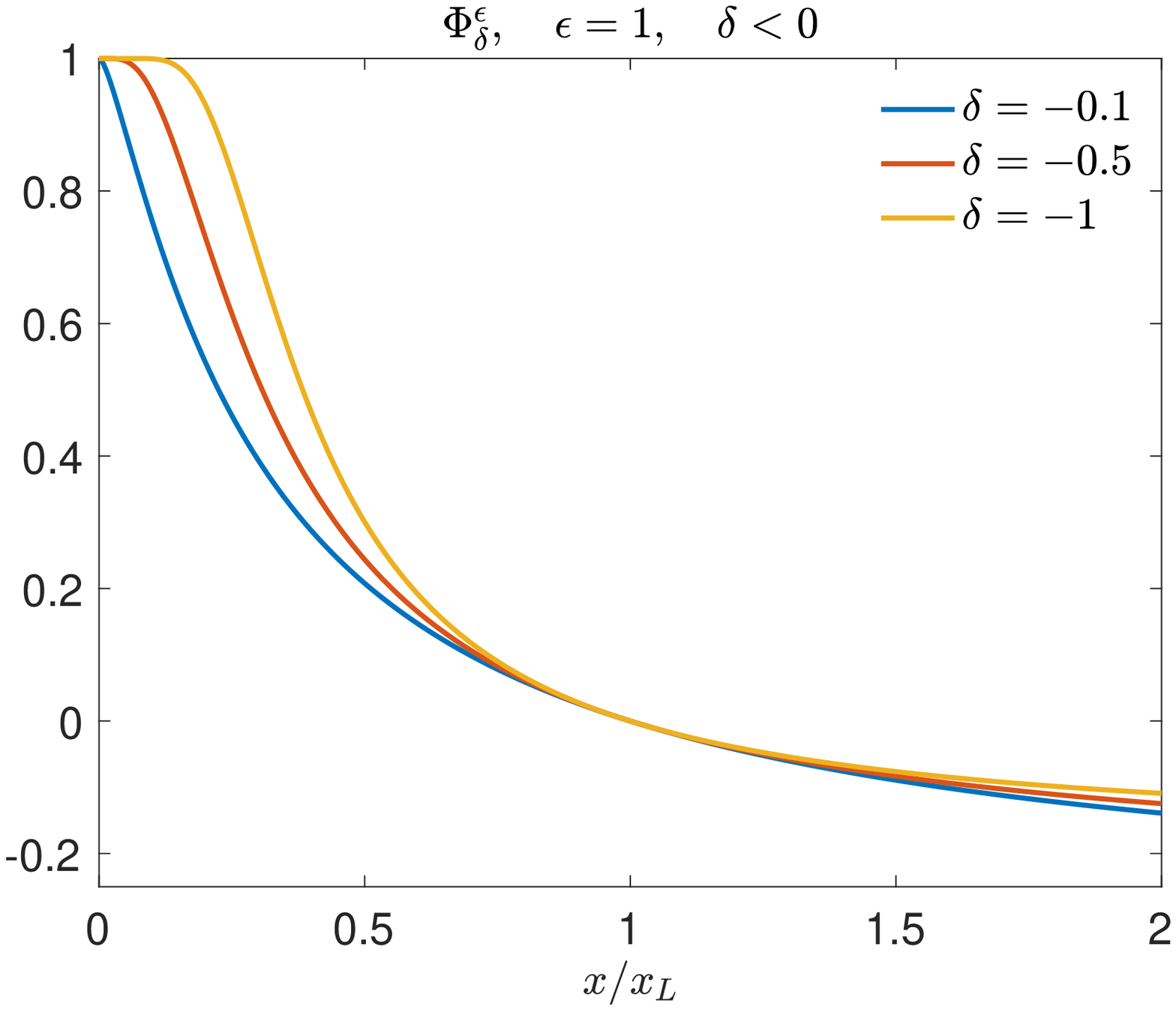}
\caption{Transition function $\Phi_\delta^\epsilon$ in \eqref{eq:Phi} for $\epsilon = 1$. In both cases we considered the choice $\mu = \lambda = \frac{1}{2}$. {  Hence, according to \eqref{eq:Phi} we have $\Phi_\delta^\epsilon \in (-1/3,1)$. We can observe the asymmetry around the value $s = x/x_L = 1$ described by the introduced transition functions in the case $\delta>0$ (left) and $\delta<0$  (right).  It is evident how  $\Phi_\delta^\epsilon$ for $\delta>0$ are increasing and convex for values $s\le 1 $ whereas, if $\delta<0$ the transition function $\Phi_\delta^\epsilon$ become concave in an interval $[0,\bar s]$, $\bar s<1$ and then convex.  }}
\label{fig:Phi}
\end{figure}

It is important to remark that  the function \fer{Phi0} is characterized by a certain asymmetry with respect to the value $s=1$, namely to the point in which the number of cancer cell reaches the carrying capacity, and  $\Phi_0^\epsilon $ vanishes. Indeed, for $0<\Delta s<1$ it holds 
\be\label{good}
\Phi_0^\epsilon(1-\Delta s) > - \Phi_0^\epsilon(1+\Delta s).
\ee
This inequality translates at the mathematical level a fundamental property: it is always easier to reach the value $s=1$ starting from below,  than to approach it from above (cf. the left case of Fig. \ref{fig:Phi}). 

Exactly in reason of inequality \fer{good}, the function $ \Phi_0^\epsilon(\cdot)$
has been recently considered in some recent work devoted to understand the reasons behind the formation of certain statistical distributions in human phenomena \cite{GT1,GT2}. In this case,
 the shape of the transition function has been designed to satisfy the main requirements of the prospect theory of Kahneman and Twersky  \cite{KT},  concerned with the description of decision under risk. 

 The transition function \fer{Phi0} is  a particular case of the general class of transition functions $\Phi^\e (\cdot)$, defined by
\begin{equation}\label{eq:Phi}
\Phi^\e(s) =\Phi^{\epsilon}_\delta(s) =  \mu \dfrac{1-e^{ \epsilon (s^\delta-1)/\delta}}{(1+\lambda) e^{ \epsilon (s^\delta-1)/\delta}+1-\lambda},
\end{equation}
where  the constant  $-1 \le \delta \le 1$,  while $0 < \mu < 1 $, and $0 \le \lambda < 1$.
Indeed, the limit case $\delta \to 0$ corresponds to the function \fer{Phi0}.

Similarly to the case of the transition function defined in \eqref{Phi0}, it can be easily verified that, for every value of the parameters $\delta, \lambda$ and $\mu$, the function $\Phi^\e(s)$ is decreasing in $s$,  equal to zero at the reference point $s=1$,  and  satisfies the bounds
 \be\label{b+}
-\frac\mu{1+\lambda}  \le \Phi_\delta^\e(s) \le \mu \frac{1 -e^{-\e/\delta} }{(1+\lambda)e^{-\e/\delta}+1-\lambda } , \quad\textrm{if}\qquad \delta >0,
  \ee
while 
\be\label{b-}
 \mu \frac{1-e^{-\e/\delta}  }{(1+\lambda)e^{-\e/\delta}+1 -\lambda}  \le \Phi_\delta^\e(s) \le   \frac\mu{1-\lambda}, \quad\textrm{if}\qquad \delta <0.
  \ee
{ Unlike the transition function \fer{Phi0}, while the parameters $\lambda$ and $\mu$, are linked to  the maximal amounts of the deterministic variations of the number $x$ in a single interaction, now the upper bound in \fer{b+} and the lower bound in \fer{b-} also depend on the parameter $\e$.  However, in all cases $\Phi_\delta^\e(s)$  still satisfies the bounds \fer{b+-}. }

A further property of the transition functions \fer{eq:Phi} is related to their dependence on the variable $\e$. With the notation $z=z(s) = (s^\delta -1)/\delta$ we have
 \be\label{gro}
 \frac{\partial \Phi_\delta^\e(s)}{\partial \e} = -2 \mu z  \frac 1 {\left[ (1+\lambda)e^{(\e z)/2} + (1-\lambda)e^{-(\e z)/2}\right]^2}.
  \ee  
Now, observing that the function
 \[
 h(y) = \left[ (1-\mu)e^{y/2} + (1+\mu)e^{-y/2}\right]^2
 \]  
 has a maximum in the point
 \[
 \bar y = \log \frac {1-\lambda}{1+\lambda},
 \]
 where
 \[
 h(\bar y) = 4(1-\lambda^2),
 \]
 we can easily determine the bound
 \[
 \left|\frac{\partial \Phi_\delta^\e(s)}{\partial \e}\right| \le \frac \mu{1-\lambda^2} \left|\frac{s^\delta -1}\delta\right|.
 \]
 This implies that, for a given $x >0$
 \be\label{okk}
  \Phi_\delta^\e\left(\frac x{x_L}\right)\, x \le \e \frac \mu{\delta(1-\lambda^2)} \left|\left(\frac x{x_L}\right)^\delta -1\right|\, x,
 \ee 
which clarifies the way in which the parameter $\e$ tunes the growth of the deterministic part of the variation \fer{eq:inter1}.

While the whole class of transition functions \fer{eq:Phi} satisfies the same type of asymmetry around the reference value $s=1$, the consequent behavior is typical of very different phenomena. Learning from the application of the transition functions \fer{eq:Phi} in the field of kinetic theory of social phenomena,   values $\delta >0$  are typical of phenomena in which the initial growth is statistically relevant. Indeed,  values $\delta >0$ have been introduced to model the statistical distribution of alcohol consumption \cite{DT}, and, more in general, the statistical distribution of addiction phenomena \cite{Tos2}. On the contrary, the case $\delta <0$ is typical of phenomena in which the initial growth is not statistically relevant, and was recently considered in \cite{DT2} to understand the formation of a social elite in consequence of the social climbing activity. }
 
Precisely,  in absence of fluctuations, the elementary interaction \fer{eq:inter1} produces a growth of the value of $x$  when $x <x_L$ for all values of the parameter $\delta$. However, in terms of $\delta$, the transition functions \fer{eq:Phi} do not behave in the same way in the region $x <x_L$, that corresponds to the interval $0 \le s \le 1$.  As remarked in \cite{DT2}  the transition functions \fer{eq:Phi} with index $\delta >0$ are increasing and convex for $s \le 1$, while the transition functions with index $\delta <0$ are concave in an interval $[0, \bar s)$, with $ \bar s <1$, and then convex, see Figure \ref{fig:Phi}. Hence, in a certain sub-interval of $[0, \bar s)$ the growth induced by the transition functions with $\delta <0$ is lower than the growth induced by the transition functions with $\delta >0$.
For this reason, the transition functions with index $\delta <0$ seem more adapted to describe the growth of cancer cells, since the presence of the inflection point in the region $x < x_L$ reflects the tendency of the body to react to the growth of cancer cells at least when their number is below a certain value. 

A second fact which leads to prefer the mechanism of growth corresponding to a transition function with $\delta <0$ is related to the behavior of $\Phi^{\epsilon}_\delta(s)$ in the interval $s >1$, namely in the interval where the number of cancer cells is above the reference value  $x_L$. In this interval the transition functions are negative and satisfy the lower bound
 \[
\Phi_\delta^\e(s) \ge \mu \frac{1-e^{-\e/\delta}  }{(1+\lambda)e^{-\e/\delta}+1 -\lambda}, 
 \]
Hence, in the interval $x>x_L$, the transition functions with $\delta <0$ take values in a small interval of size approximately $\e/|\delta|$, which corresponds, since $\e\ll 1$, to an almost negligible variation  of the deterministic part of the size, and consequently to  an effective stabilization of the size around the value $x_L$. Clearly, this property does not hold when $\delta >0$, since in this case the lower bound in \fer{b+} does not depend on $\e$.

Once the deterministic mechanism of growth has been quantified in terms of the transition functions \fer{eq:Phi}, the upper bound in \fer{b+-} allows to compute the lower bound relative to the random fluctuations which can be consistently inserted into the elementary interaction \fer{eq:inter1} to preserve the positivity of the variable $x$. Indeed
$ x^\prime \ge 0$ independently of $\e$ if
 \be\label{b-eta}
 \eta_\e \ge -1 + \frac\mu{1 + \lambda}.
 \ee 
   { It is important to remark that the whole class of transition functions defined in \fer{eq:Phi} satisfy condition \fer{scala}. Indeed, in the limit $\e \rightarrow 0^+$ the transition function satisfies
 \[
\Phi^{\e}(s) \approx \mu \e  \dfrac{(s^\delta-1)/\delta}{(1+\lambda)\epsilon (1-s^\delta)/\delta + 2},
\]
 which implies 
\be\label{drift}
\lim_{\epsilon \to 0^+} \frac{\Phi^{\e}(s)}\epsilon = \frac\mu{2\delta} (1-s^\delta)
\ee
Hence, if in the elementary interaction we consider a transition function in the class  \fer{eq:Phi}, for any given $-1 \le \delta \le 1$ the kinetic equation \fer{eq:boltz_1} in the quasi-invariant limit is given by the Fokker--Planck equation 
\begin{equation}\label{eq:FP1}
\partial_{t} f(x,t) = \partial_x \left[\dfrac{\mu}{2\delta} \left( \left( \dfrac{x}{x_L} \right)^\delta -1\right) x f(x,{t}) +\dfrac{\sigma^2}{2} \partial_x (x^2 f(x,{t}))\right].
\end{equation}
We notice that the drift term in \eqref{eq:FP1} takes the form of classical growth laws recalled for the convenience of the reader in Appendix \ref{app:A}. For instance, $\delta = 1$ corresponds to logistic growth, $\delta\rightarrow 0$ to Gompertzian growth, and $\delta<0$ to von Bertalanffy-type growth law. 
}
\subsection{ Relevant cases}

The Fokker--Planck equation \fer{eq:FP1} retains memory of the kinetic description through the relevant parameters of the transition function \fer{eq:Phi}, namely the parameters $\delta$ and $\mu$, and through the shape of the drift term, as given by \fer{drift}. However, the parameter $\lambda$ is lost in the limit. Also, the details of the variable $\eta_\e$ are lost in the limit passage, so that the role of fluctuations is taken into account only through their variance, parameterized by $\sigma$.  As we shall see, at difference with the others, the  value of the parameter $\delta$ fully characterizes the shape of the steady state of equation \fer{eq:FP1}. 

A distinguished case is obtained by taking $\delta \to 0$ in Eq. \fer{eq:FP1}. The resulting Fokker--Planck equation in this case is given by
\[
\partial_{t} f(x,t) = \partial_x \left[ \frac{\mu}{2} \log \frac x{x_L} x f(x,{t}) +\dfrac{\sigma^2}{2} \partial_x (x^2 f(x,{t}))\right],
\]
which is the equation considered in \cite{AG}.   {This allows to compare out next results on the controlled Fokker--Planck equation with the results obtained in \cite{AG}. As usual for Fokker--Planck type equations in divergence form, it is immediate to evaluate the explicit shape of its stationary distribution, and to clarify, resorting to this shape, the consequences of the different transition functions on the statistical tumor growth}.

\begin{figure}
\centering
\includegraphics[scale = 0.35]{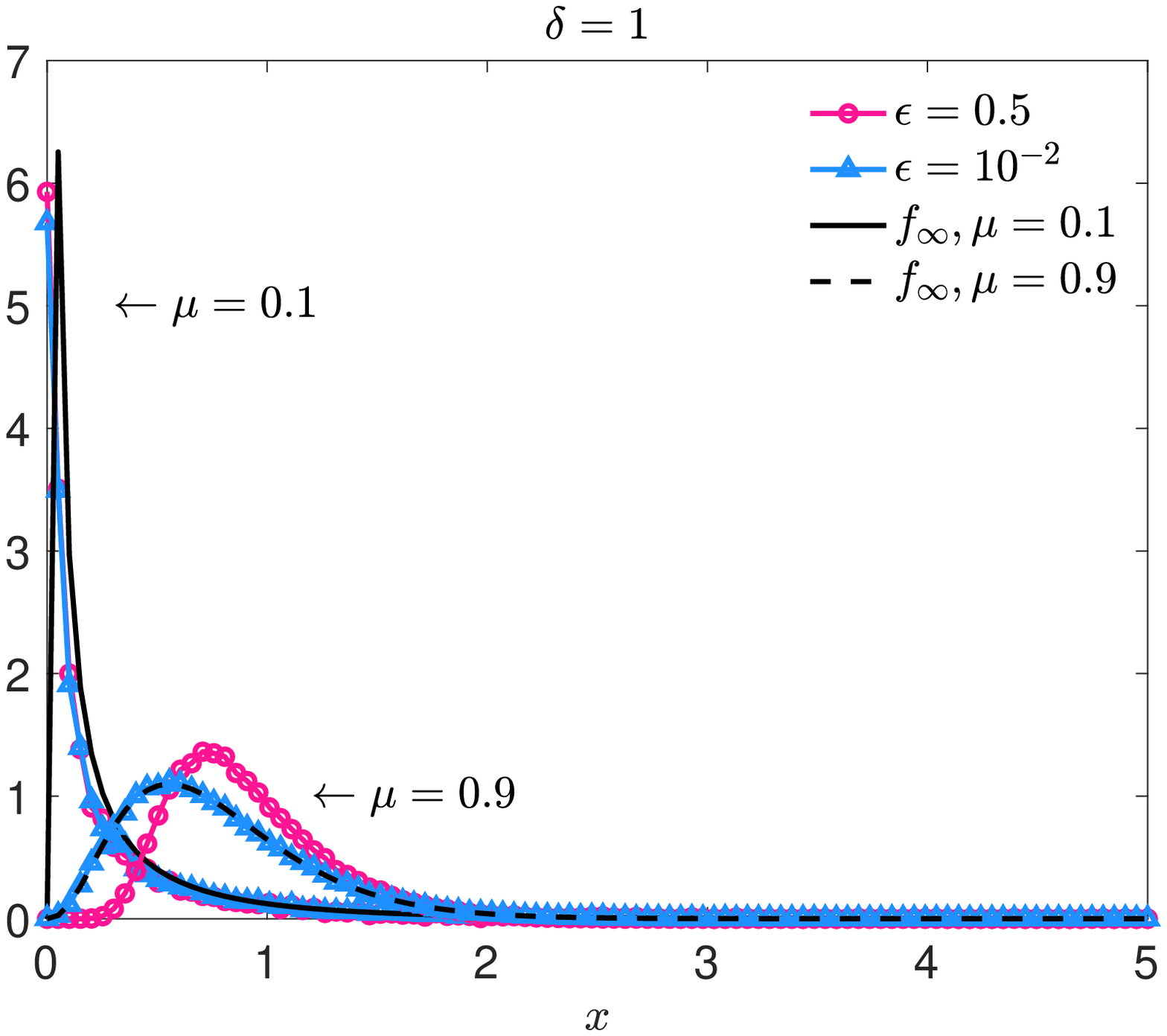}
\includegraphics[scale = 0.35]{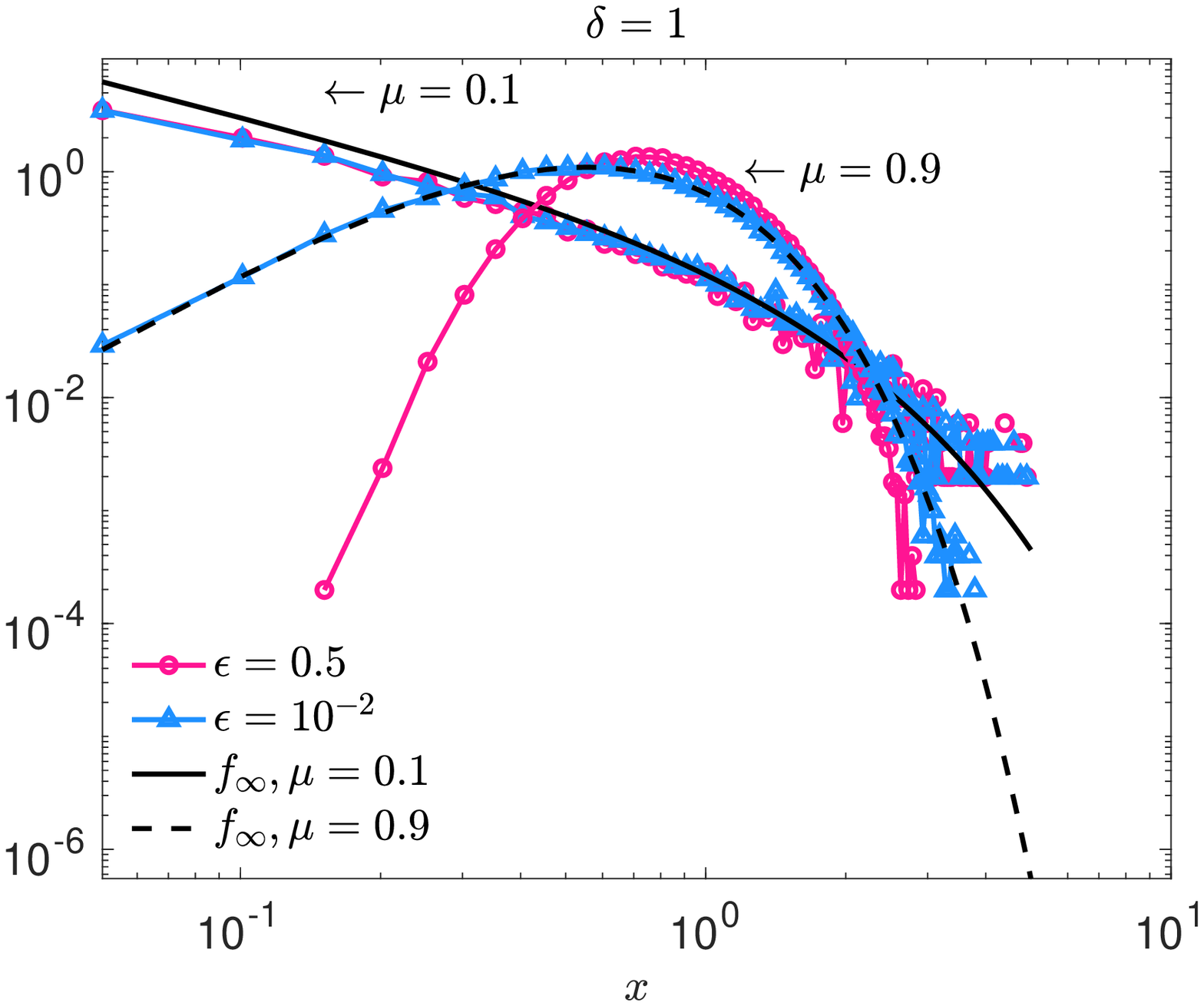}\\
\includegraphics[scale = 0.35]{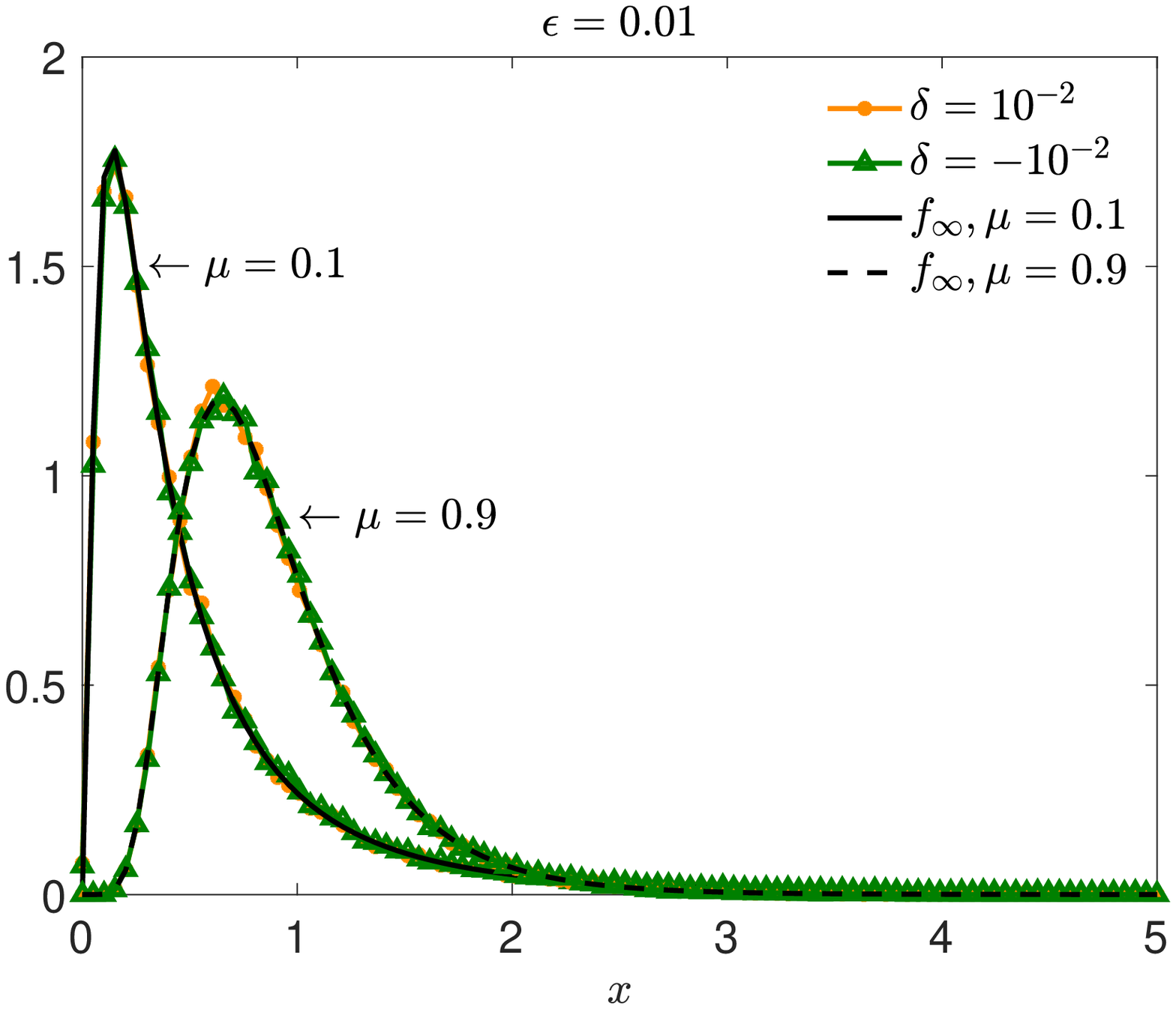}
\includegraphics[scale = 0.35]{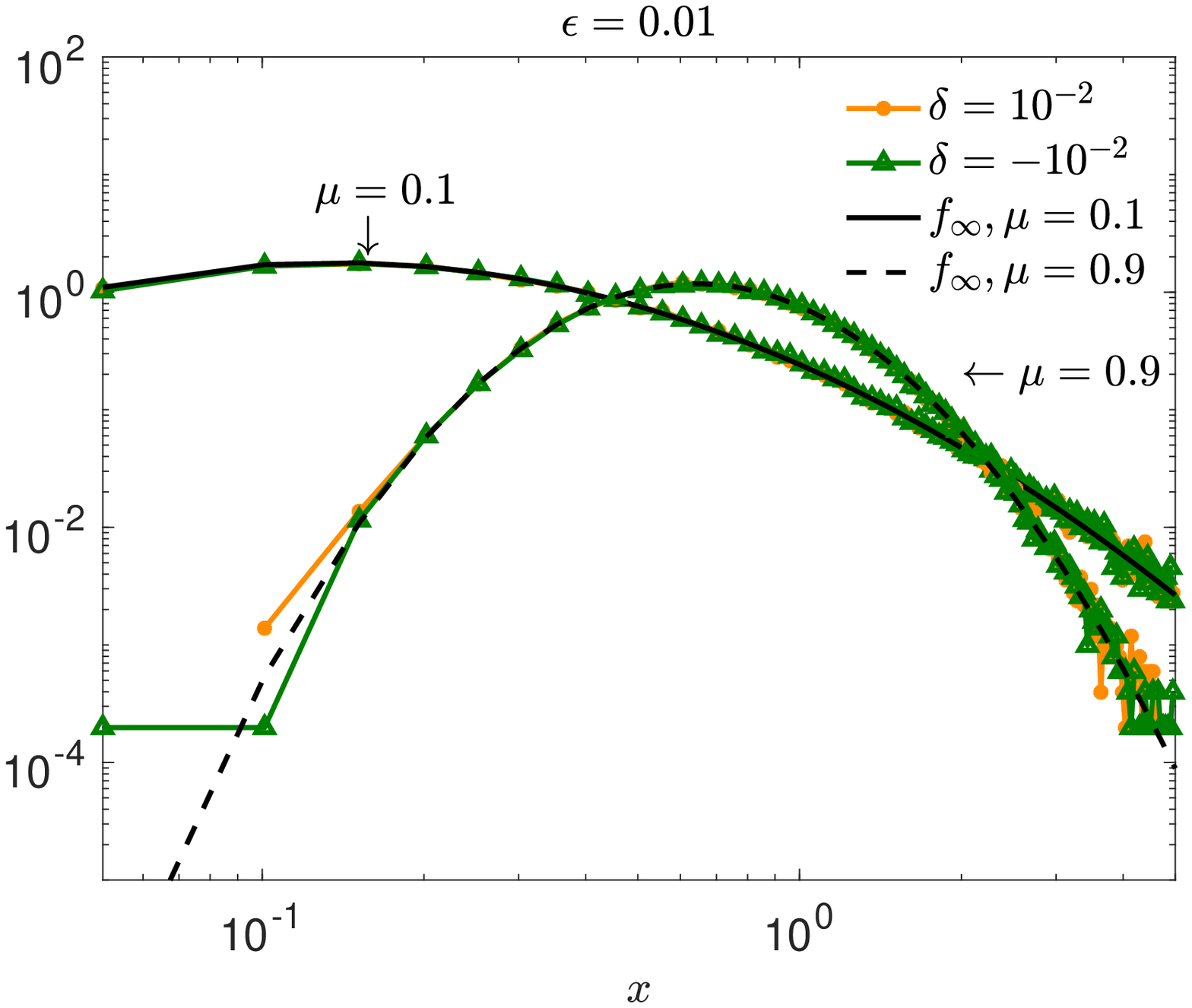} \\
\includegraphics[scale = 0.35]{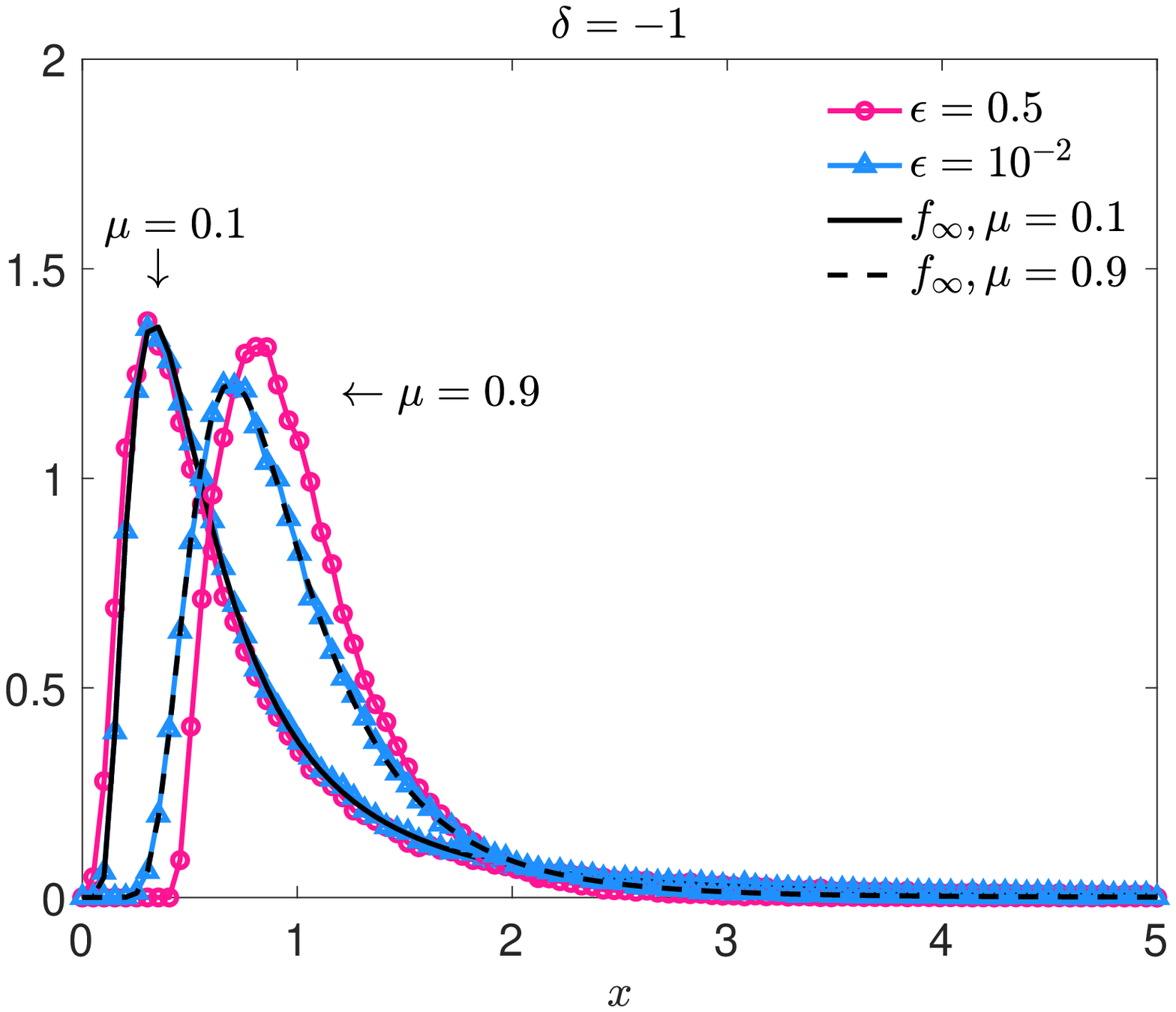}
\includegraphics[scale = 0.35]{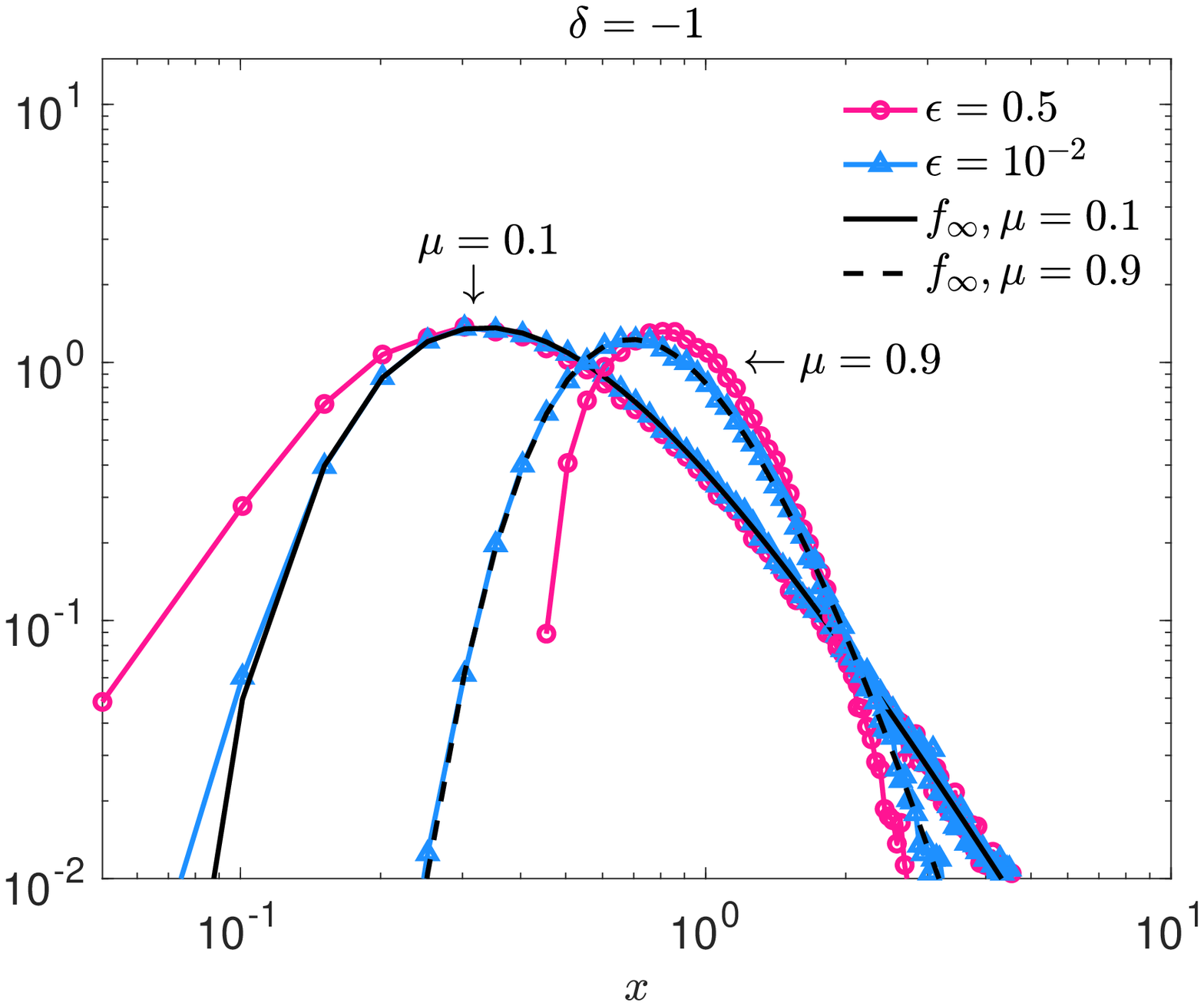}
\caption{Comparison of the analytical steady states $f_\infty$ (full and dashed curves) given in \eqref{gg}-\eqref{aa}-\eqref{eq:lognormal} (from top to bottom) with the numerical solution of the Boltzmann-type equation \eqref{eq:boltz_1}  for large times in the quasi-invariant regime for $\epsilon\ll1$ (marked curves). We considered $\delta = 1$ (top row), $\delta \approx 0$ (middle row), and $\delta = -1$ (bottom row). It is easily observed how we consistently catch the obtained equilibrium distribution in all regimes. In the figures on the right we highlight the tail behavior  plotting the distribution in loglog scale for all  the considered regimes. In all the reported numerical results we considered $\lambda = \mu =0.1,0.9$, $x_L = 1$, and $\sigma^2 = 0.2$.} %{\bf Perch\'e alcune curve a destra si fermano prima di toccare gli assi (per bassi $x$)?} }
\label{fig:equilibrium}
\end{figure}
 
 \subsection{Steady states}

Let $\gamma = \mu/\sigma^2$, and suppose that $\gamma >\delta$. Note that this condition is restrictive only when $\delta >0$. Then, the  asymptotic distribution $f_\infty(x)$  satisfies the first order differential equation
\[
\partial_x (x^2 f(x,{t})) + \dfrac{\gamma}{\delta}\left(\left( \dfrac{x}{x_L} \right)^\delta-1 \right)x f(x,{t}) = 0, 
\]
whose solution is given by
\be\label{equilibrio}
f_\infty(x) = f_\infty(x_L) \left( \dfrac{x}{x_L} \right)^{{\gamma}/{\delta}-2} \exp\left\{ - \dfrac{\gamma}{\delta^2} \left( \left( \dfrac{x}{x_L} \right)^\delta -1 \right) \right\}, 
\ee
see also \cite{DT}. It seems worthwhile to remark that the obtained equilibrium distribution \eqref{equilibrio} lies on  wider classes of probability distributions depending on the sign of the parameter $\delta$. 

\paragraph{The case $\delta>0$.}

Let us fix the mass of the steady state \fer{equilibrio} equal to one. If $\delta >0$, the consequent probability density is a generalized Gamma. These distributions are characterized in terms of a shape $\kappa>0$, a scale parameter $\theta >0$, and the exponent $\delta >0$. Therefore \eqref{equilibrio} can be parametrized in terms of the introduced parameters as follows
 \be\label{gg}
f_{\infty}^{\kappa,\delta,\theta}(x) = \frac\delta{\theta^\kappa \Gamma \left( \kappa/\delta\right)}\, x^{\kappa -1} \exp\left\{ -\left( \frac x\theta\right)^\delta \right\},
 \ee
where shape and the scale parameter of the equilibrium state are given by  
 \[
 \kappa =  \frac\gamma\delta -1,  \quad  \theta =  x_L \left( \frac{\delta^2}\gamma\right)^{1/\delta}. 
 \]
 We point the interested reader to \cite{Lie,Sta} for further details.

 \begin{remark}
The logistic growth \fer{logi} corresponds to the value $\delta =1$. In this case, the steady state of the corresponding Fokker--Planck equation is a Gamma density function, with exponent $\kappa = \gamma -1$ and scale parameter $\theta = x_L/\gamma$. The condition $\kappa >0$ is satisfied if $\mu > \sigma^2$, that is when the elementary transition \fer{eq:inter1} is characterized by random fluctuations that are small with respect to the deterministic part. The values of $\delta <1$ correspond to generalized logistic growth laws, as given by \fer{gen}. In all cases, the consequent generalized Gamma densities decay to zero exponentially {  as $x\to +\infty$}.  
\end{remark}

\paragraph{The case $\delta<0$.}

 In the case of negative $\delta$'s, corresponding to the introduced von Bertalanffy growth \fer{eq:vonB}, we notice a different behavior for large values of $x$.  
Indeed, the equilibrium distribution \eqref{equilibrio} is an Amoroso-type, or power law distribution \cite{Amo}
\be\label{aa}
f_\infty^{\kappa,|\delta|,\theta}(x) = \frac{|\delta|}{ \Gamma \left( \kappa/|\delta|\right)}\,\frac {\theta^\kappa}{x^{\kappa + 1}} \exp\left\{ -\left( \frac\theta{x}\right)^{|\delta|} \right\},
 \ee
which is characterized by a polynomial decay. The shape and the scale parameter of the equilibrium state \fer{equilibrio} are given by  
\be\label{para2}
 \kappa =  \frac{\gamma}{|\delta|} + 1,  \quad  \theta =  x_L \left( \frac\gamma{\delta^2}\right)^{1/|\delta|}.
 \ee
In reason of the polynomial decay of \fer{aa}, the equilibrium density has moments bounded only of order $p < \kappa$. The case $\delta = -1$ corresponds to the inverse Gamma distribution.   {As documented in \cite{Gab,New,PT2} through an exhaustive list of references,  power-law distributions occur in an extraordinarily diverse range of situations}. 
  {In biology, power laws have been claimed to describe the distributions of the connections of enzymes and metabolites in metabolic networks, the number of interactions partners of a given protein, and other quantities \cite{Karev, Kuz}. }
In the present context, these distributions, characterized by polynomially-decaying tails,  indicate higher probabilities of having tumors with a big size. Therefore, the paramount need of identifying therapeutical protocols aimed at reducing the probability of having big tumors translates from a statistical point of view in dampening the mass of the tails. 

\paragraph{The case $\delta \to 0$.}
The limit case $\delta \to 0$ corresponds to Gompertz growth \eqref{eq:gomp}. The equilibrium density is easily seen to be the lognormal equilibrium
\begin{equation}
\label{eq:lognormal}
f_\infty(x) = \dfrac{1}{\sqrt{2\pi\gamma} x} \exp\left\{ -\dfrac{(\log x-\kappa)^2}{2\gamma} \right\}, 
\end{equation}
where  $\kappa = \log x_L - \gamma$. This border case still corresponds to a density function with slim tails. \\

In Figure \ref{fig:equilibrium} we represent the numerical approximation of the Boltzmann-type model \eqref{eq:boltz_1} in the quasi-invariant regime through Direct Stochastic Monte Carlo (DSMC) methods, see \cite{PR,PT2} for an introduction. In details, we considered the initial distribution 
\begin{equation}
\label{eq:f0}
f(x,0) = 
\begin{cases}
1 &  x\in[1,2]\\
0 & \textrm{elsewhere},
\end{cases}
\end{equation}
and $N = 10^5$ particles. Furthermore, we considered the following choice of parameters $\mu = 0.1,0.9$, $x_L = 1$ and $\sigma^2 = 0.2$. It can be easily observed how the reconstructed large time distribution from the Boltzmann model can be approximated with the steady state of the Fokker-Planck models, producing therefore the correct tails of the various equilibrium distributions.

\section{The controlled model}\label{control}

In Section \ref{kinetic} we {  introduced and discussed a variety of kinetic models} suitable to describe tumor growth.  The main brick of this construction was the choice  of  the class of transition functions \fer{eq:Phi} entering the elementary interaction \fer{eq:inter1}, and characterizing the growth in terms of the parameter $\delta$ ranging from $-1$ to $+1$. In particular, it was shown that, for negative values of the parameter $\delta$, corresponding to von Bertalanffy growth as explained in Appendix \ref{app:B}, the resulting equilibrium in the limit of grazing interactions is given by a probability density with polynomial tails, in the form of Amoroso distribution \fer{aa}. In details, we studied how, for some values of the parameter $\delta$, the kinetic modeling  of Section \ref{kinetic}  allows to obtain Fokker--Planck type equations previously considered in the literature, even if derived in a different way.  In this direction we mention the limit $\delta \to 0$ in the introduced kinetic modeling, corresponding to Gompertz growth \cite{AG}, which exhibits lognormal equilibria \fer{eq:lognormal}. 

The new kinetic description allows to enlighten the effects of therapies by acting on the elementary responses to environmental cues directly, to show how these therapies act on the resulting Fokker--Planck equations, and ultimately to compare  the results in \cite{AG} with the present ones. In this direction, we will consider a therapy like a control acting on the elementary transitions to minimize the growth. 
%In the following we study two possible therapies 

%\subsection{Constrained elementary interactions}

To study the effect of therapies on the growth process we consider a constrained version of the transition model \eqref{eq:inter1} which depends on a control $u$ representing the instantaneous correction due to an external action. This control can be additive  
\begin{equation}\label{eq:additive}
x^\prime  = x + \Phi^\e(x/x_L)x + \e x\,u + x\eta_\e, 
\end{equation}
and in this case the effect of $u$ is to modify \emph{at best} the growth in an additive way, or multiplicative
\begin{equation}\label{eq:inter_cons}
x^\prime  = x + u\Phi^\e(x/x_L)x + x\eta_\e, 
\end{equation}
which implies a direct action on the transition function. The former will be discussed in Section \ref{addi} and the latter in Section \ref{multi}. 

In both cases the control variable is given by a multiplicative coefficient of the variable $x$, meaning that the control acts similarly on single cells, so that the eventual control is proportional to tumor size. Furthermore, we observe that a control of the form \eqref{eq:additive} induces an external modification of the death rate. On the other hand the multiplicative control of the form \eqref{eq:inter_cons} modifies directly the dynamics acting on the balance between death and birth. Moreover, in the additive control, the size of the controlled variable is tuned by the small parameter $\e \ll 1$. 
%We start by studying the additive control given by \fer{eq:additive}, leaving to the next Section the  multiplicative   case  \fer{eq:inter_cons}

The optimal control $u^*$ {  can be} determined as the minimizer of a cost functional 
\begin{equation}\label{eq:optimal_u}
u^* = \textrm{arg}\min_{u\in \mathcal U} \dfrac{1}{2} J(x^\prime, u),
\end{equation}
subject to the constraint \eqref{eq:additive}. In \fer{eq:optimal_u} the minimum is taken on the space $\mathcal U$ of all admissible controls. In the following we will consider a quadratic cost functional in the form
\be\label{quadra}
J(x^\prime , u)= \dfrac{1}{2} \left\langle (x^\prime - x_d)^2 + \nu_\e u^2 \right \rangle,
\ee
being $\nu_\e >0$ a penalization coefficient and $x_d>0$ is the desired tumors' size that one would like to reach.   {According to \eqref{quadra}, the cost increases quadratically with the distance to a desired size, this choice mimics the fact that more efforts are needed to contain tumours with bigger size, i.e. cancers that are detected too late.} This could be different than zero allowing the existence of tumors with a controlled size. The presence in \fer{quadra} of the mean operator $\langle \cdot\rangle$ permits to obtain a control which does not depend on the presence of the random fluctuations. 
The goal  is to obtain a control which minimizes the distance with respect to the desired size $x_d\in \mathbb R_+$. The minimization of  \eqref{eq:optimal_u} can be classically done resorting to a Lagrange multiplier approach        {in the present setting. It is worth to observe that other convex cost functionals may be considered leading often to problems whose analytical solution cannot be obtained explicitly. Therefore, in more general settings suitable numerical methods should be developed, see e.g. \cite{AFK}. }

\subsection{Additive control and equilibrium distribution}\label{addi}
We concentrate first on a dynamics embedded with an additive control strategy \eqref{eq:additive} seeking to minimize the cost functional \eqref{eq:optimal_u}. Hence, we consider the Lagrangian
\[
\mathcal L(u,x^\prime) = J(x^\prime,u) + \alpha \left \langle x^\prime - x -\Phi^\e(x/x_L)x -\e\, x u  - x\eta_\e \right \rangle,
\]
where $\alpha \in \mathbb R$ is the Lagrange multiplier associated to the constraint \eqref{eq:additive}. The optimality conditions read
\[
\begin{cases}
\partial_u \mathcal L(x^\prime, u) =  \nu_\e u - \alpha \e\, x = 0\\
\partial_{x^\prime} \mathcal L(x^\prime, u) = \left \langle x^\prime - x_d \right \rangle + \alpha = 0.
\end{cases}
\]
Eliminating the Lagrange multiplier yields the optimal value
\be\label{op-ad}
 u_* = - \dfrac{\e\,x}{\nu_\e + \e^2x^2 } \left( x - x_d + \Phi^\e(x/x_L)x\right). 
\ee
Plugging the optimal value \fer{op-ad} into \fer{eq:additive} gives the following optimal constrained interaction 
\begin{equation}\label{eq:add}
x_*^\prime  = x + \frac{\nu_\e}{\nu_\e + \e^2 x^2}\Phi^\e(x/x_L)x - \frac{\e^2 x^2}{\nu_\e + \e^2 x^2}(x-x_d) + x\eta_\e. 
\end{equation}
Note that for $x \le x_L$ the transition function $\Phi^\e$ is nonnegative, so that the post-interaction value $x^\prime_*$  is nonnegative if the fluctuation variable $\eta_\e$ satisfies the condition
 \[
  \eta_\e \ge -1 +  \frac{\e^2 x_L^2}{\nu_\e + \e^2 x_L^2}.
  \]
In view of condition \fer{b-eta}, this condition is satisfied for $\e$ sufficiently small. 

In presence of the controlled interaction \fer{eq:add}, one can consider as before the limit of grazing interactions, provided all quantities in \fer{eq:add} scale in the right way with respect to $\e$. To this extent, it is enough to scale the penalization $\nu_\e = \e \nu$, where $\nu >0$, to get
 \be\label{scal2}
  \frac{\nu_\e}{\nu_\e + \e^2 x^2} =  \frac{\nu}{\nu + \e x^2},\qquad \frac{\e^2 x^2}{\nu_\e + \e^2 x^2} = \e \, \frac{ x^2}{\nu + \e x^2}.
  \ee
At this point, proceeding as in Section \ref{grazing} with the new elementary interaction \fer{eq:add} we obtain that the controlled kinetic model converges, in the grazing limit $\e \to 0$ to a Fokker--Planck type equation with a modified drift term, that takes into account the presence of the control. In terms of the controlled density $f_a(x,t)$, this equation reads
 \[
 \partial_{t} f_a(x,t) = \partial_x \left\{ \left[ \dfrac{\mu}{2\delta} \left( \left( \dfrac{x}{x_L} \right)^\delta -1\right) x  + \frac{x^2}\nu(x-x_d) \right]f_a(x,{t}) +\dfrac{\sigma}{2} \partial_x (x^2 f_a(x,{t})) \right\}.
 \]
\\
%\subsubsection{Equilibrium distribution for additive control}
We will refer here to the case in which $\delta <0$, which in the uncontrolled case leads to steady states with polynomial tails.  Then, in presence of the additive control, the  asymptotic distribution $f_{a,\infty}(x)$  satisfies the first order differential equation
\[
\partial_x (x^2 f_{a,\infty}(x,{t})) + \left[\frac\gamma{|\delta|} \left(1 - \left( \dfrac{x_L}x \right)^{|\delta|} \right)x  +\frac{2\,x^2}{\sigma\nu}(x-x_d)   \right]f_{a,\infty}(x,{t}) = 0, 
\]
where $\gamma = \mu/\sigma$. The solution is given by
\be\label{equi-cont}
f_{a,\infty}(x) = C(x_L,x_d) \left( \dfrac{x_L}x \right)^{{\gamma}/{|\delta|} +2} \exp\left\{ - \dfrac{\gamma}{\delta^2} \left( \left( \dfrac{x_L}x \right)^{|\delta|} -1 \right) \right\} \exp\left\{ - \frac{(x-x_d)^2}{\sigma\nu}\right\}. 
\ee
In \fer{equi-cont} the constant $C(x_L,x_d)$ is chosen such as the mass of the density function equal to one.

\begin{figure}
\centering
\includegraphics[scale = 0.35]{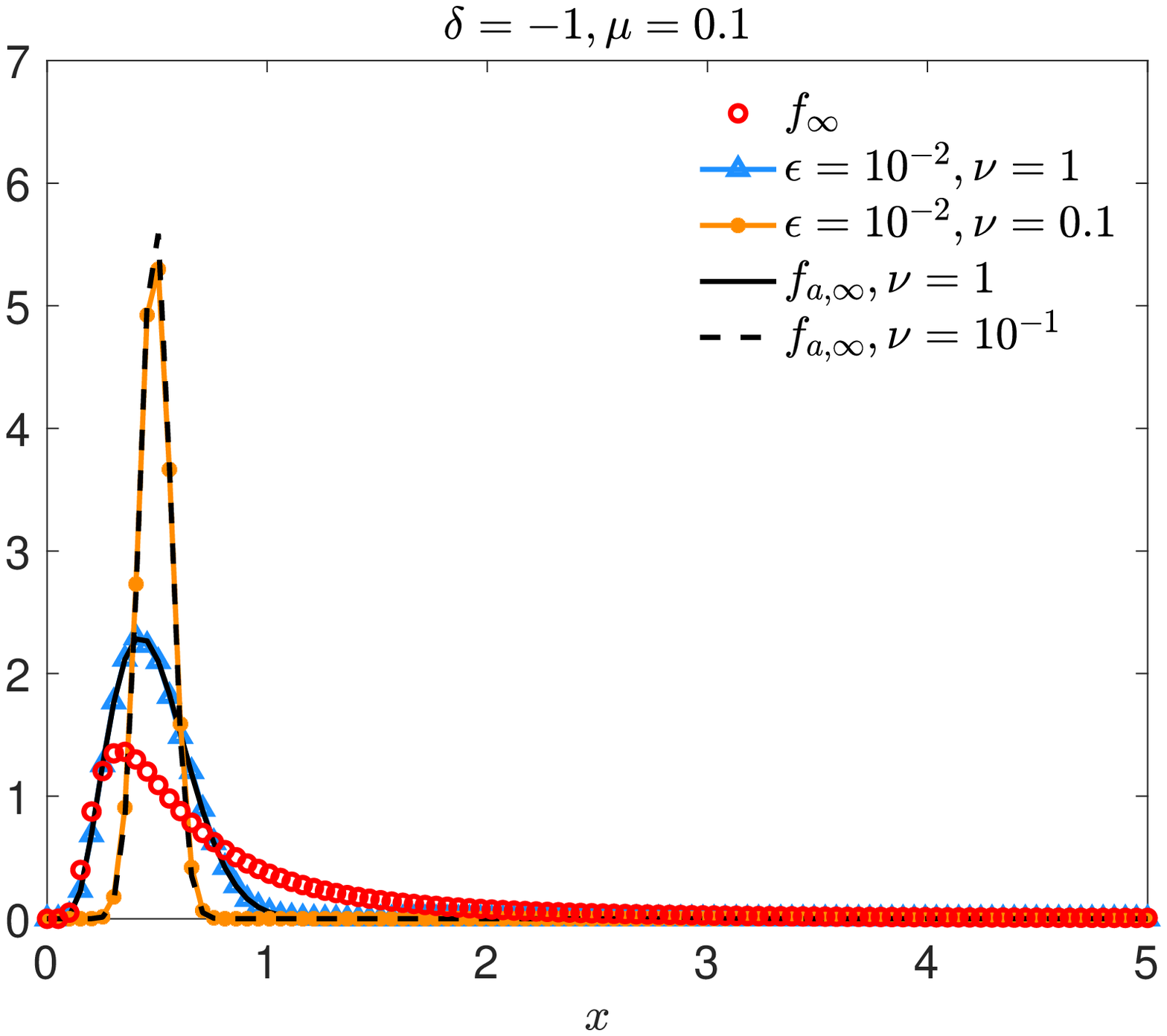}
\includegraphics[scale = 0.35]{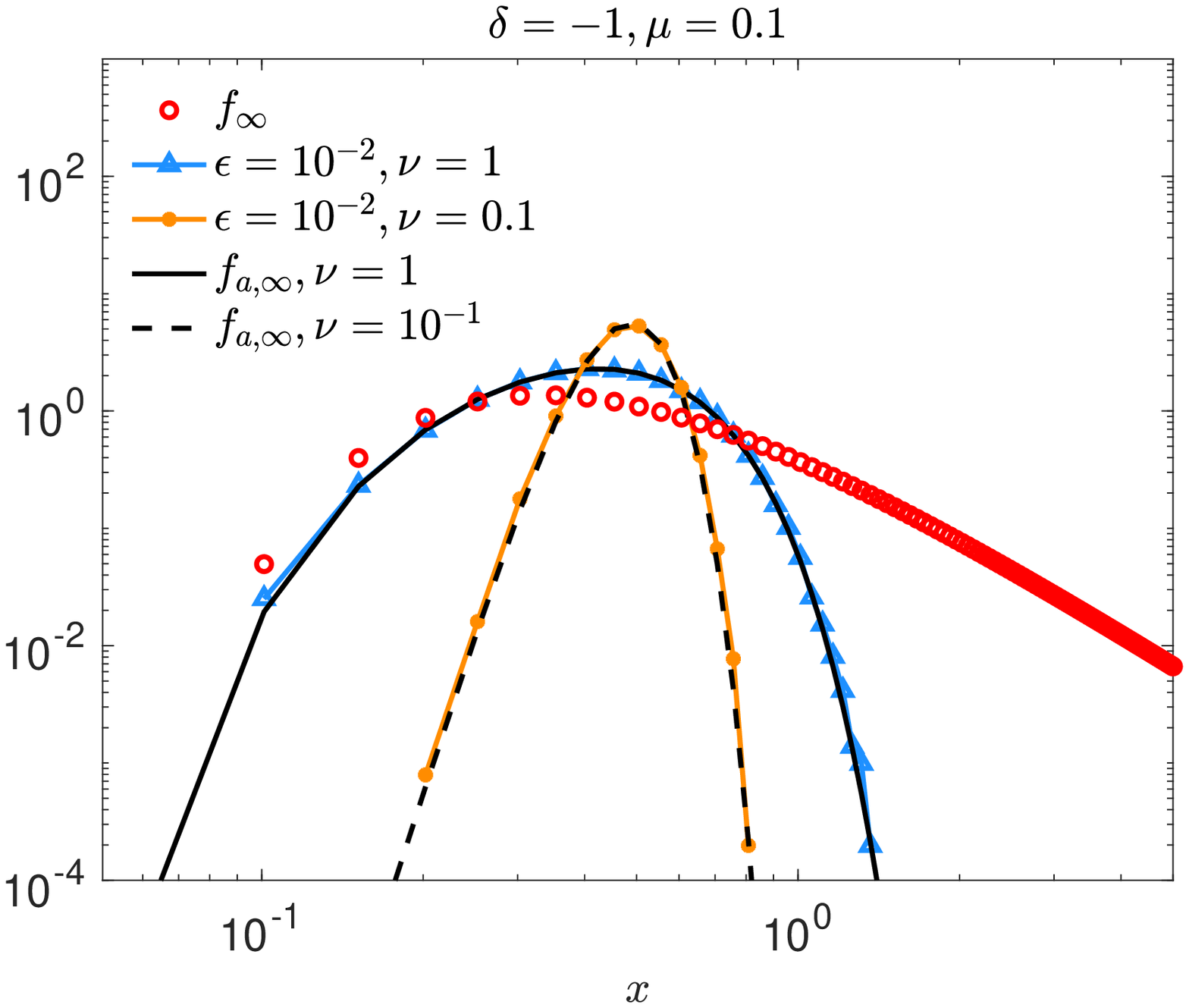}\\
\includegraphics[scale = 0.35]{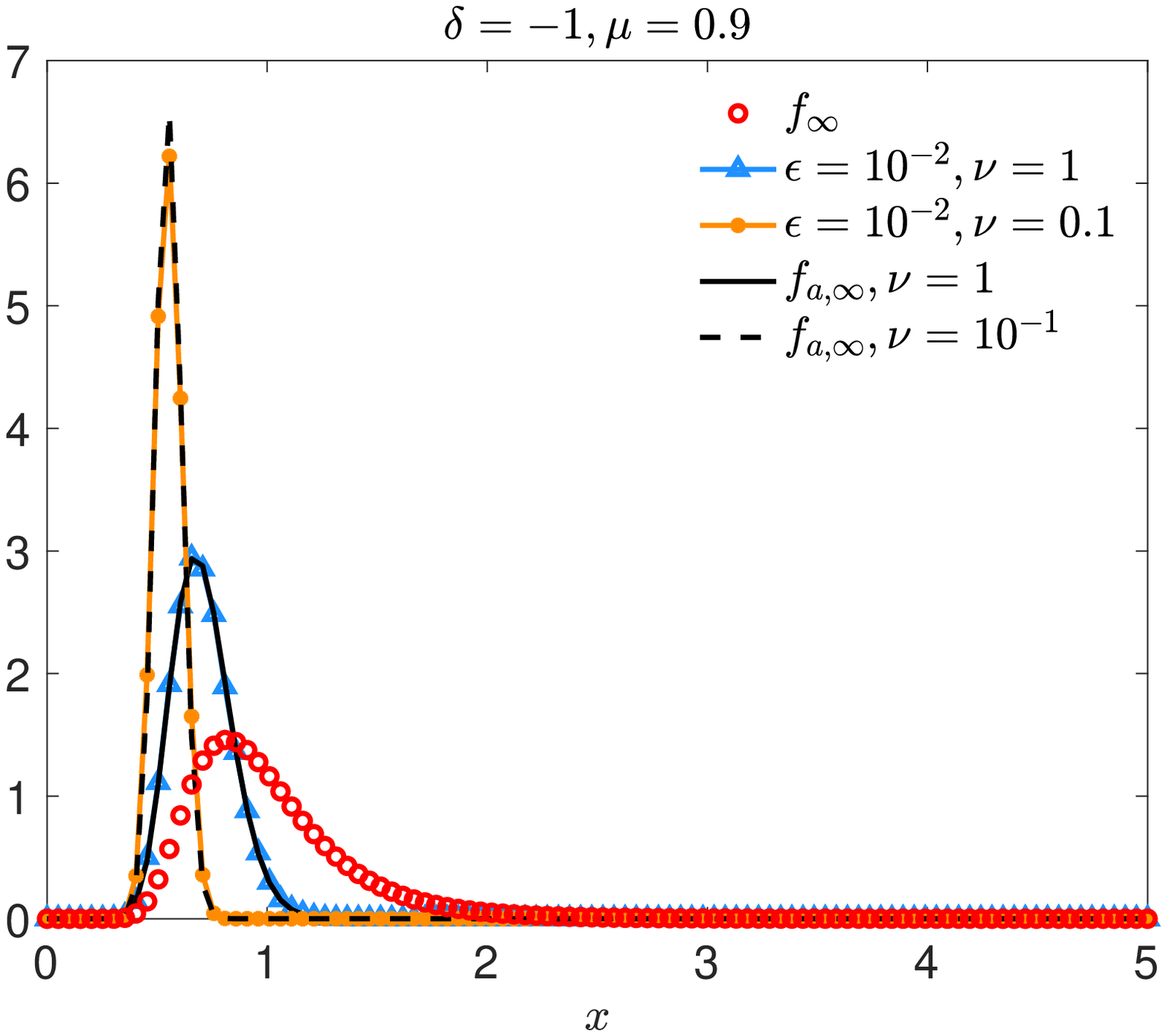}
\includegraphics[scale = 0.35]{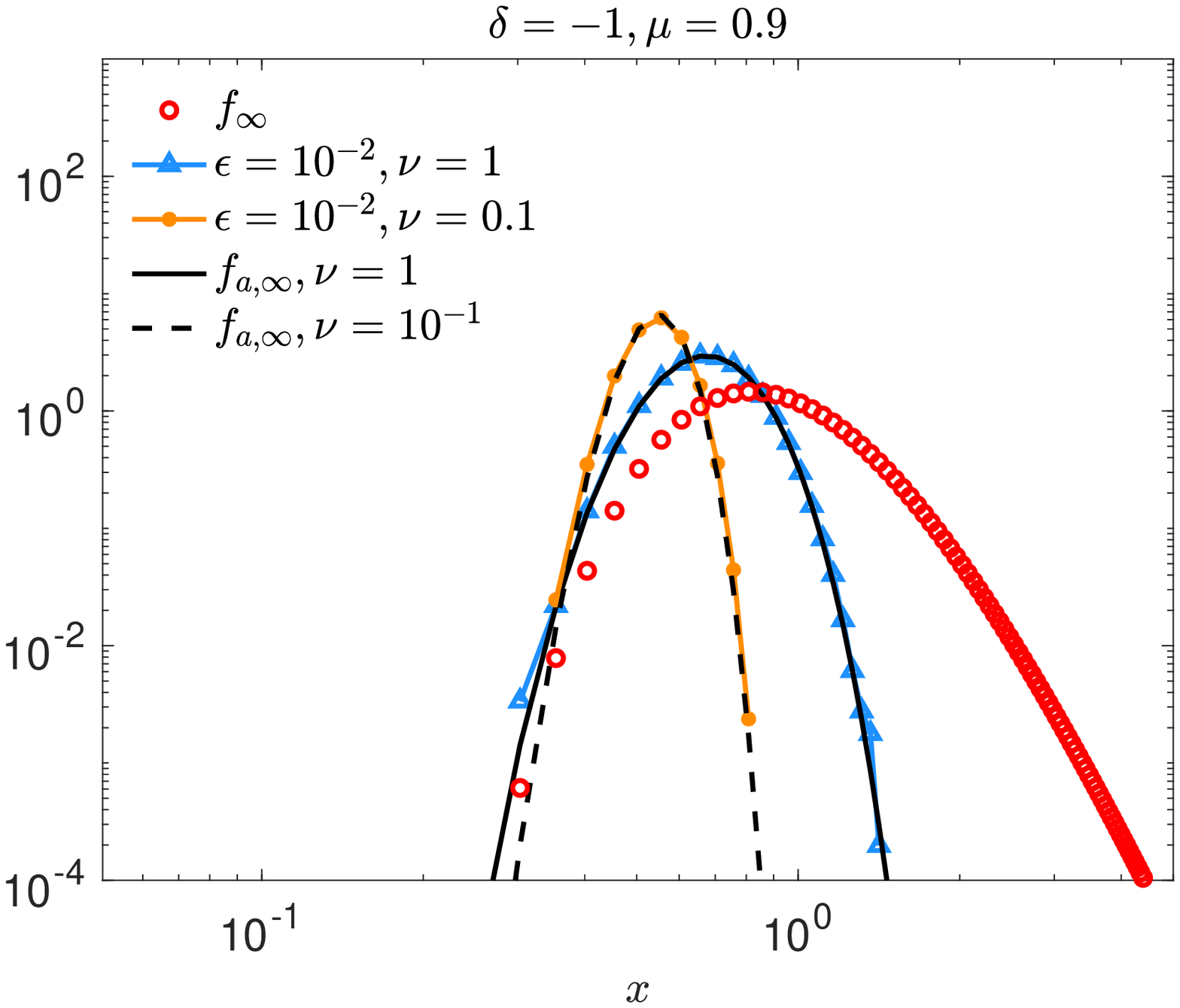}\\
\caption{Comparison of the analytical steady states \eqref{prod} with the numerical large time solution of the Boltzmann-type equation with additive constrained interaction \eqref{eq:additive} in the quasi-invariant regime for $\epsilon= 10^{-2}$ and $\nu =10^{-1},1$ in the case $\lambda = \mu=0.1$ (top tow) and $\lambda = \mu=0.9$ (bottom row). We considered $x_L = 1$ and target state $x_d =0.5$. In the right column we report the obtained distributions in loglog scale to highlight the behavior of the tails. In red dotted we report the equilibrium distribution of the unconstrained case for $\mu = 0.1$ (top row) and $\mu = 0.9$ (bottom row). }
\label{fig:additive}
\end{figure}

The steady state \fer{equi-cont} can be rewritten as the product of two probability densities. The first one is the solution of the uncontrolled Fokker--Planck equation \fer{eq:FP1}, given by the Amoroso type density \fer{aa}, with parameters 
$\kappa$ and $\theta$ in \fer{para2}.  The second term has the form of the Gaussian density 
\be\label{gauss}
\mathcal N(x_d, \sigma\nu/2)= \frac 1{\sqrt{\pi\sigma\nu}} \exp\left\{ - \frac{(x-x_d)^2}{\sigma\nu}\right\},
 \ee
of mean value $x_d$ and variance $\sigma\nu/2$. Clearly, since $x$ in \fer{gauss} ranges on the whole real line $\R$, identity holds provided the product of the densities is multiplied by the characteristic  function of the set $x \ge 0$, we denote by $\mathcal I(x \ge 0)$. Finally
 \be\label{prod}
 f_{a,\infty}(x) =  \tilde C(x_L,x_d,\sigma,\nu)\;  f_\infty^{\kappa,|\delta|,\theta}\left(x\right) \mathcal N(x_d, \sigma\nu/2)\mathcal I(x \ge 0).
 \ee
In \fer{prod} the constant $\tilde C(x_L,x_d,\sigma,\nu)>0$ is such that the density $f_{a,\infty}$ is normalized to one. It is remarkable that, at variance with the uncontrolled case,  the presence of the Gaussian density is such that the controlled distribution possesses exponentially decaying tails at infinity. Moreover, for small values of the penalization variable $\nu$, the mean value of the controlled case is close to the target value $x_d$, and the equilibrium solution has a small variance. In other words, the controlled case is such that the target value $x_d$ can substantially be reached.

In Figure \ref{fig:additive} we compare the numerical solution of the Boltzmann-type model \eqref{eq:boltz2} for large times with additive constrained transitions \eqref{eq:add} in the case $\epsilon = 10^{-2}$. In details, we considered the case $\delta = -1$, the initial distribution \eqref{eq:f0} and the scaled penalization $\nu = 10^{-1},10^0$. Here, we supposed that the target size is $x_d = \frac{1}{2}$ whereas $x_L = 1$. We can observe how the numerical large time distribution is consistently described by the derived equilibrium distribution of the Fokker-Planck model \eqref{equi-cont} for sufficiently small $\epsilon \ll 1$. It is easily observed how for decreasing penalizations the equilibrium distribution $f_{a,\infty}$ tends to concentrate around the target size $x_d$ with decreasing variance coherently with what we obtained in \eqref{prod}. The effect of the control on the tails of the distribution is highlighted by direct comparison with the equilibrium distribution of the unconstrained case of the form \eqref{aa}.

\begin{remark}
We can observe how the introduced control needs to modify the growth term to influence the behavior of the tails of the emerging equilibrium distribution. Indeed, if we consider a control that minimizes the cost \eqref{eq:optimal_u}-\eqref{quadra} subject to the following dynamics 
\[
x^\prime = x + \Phi^\epsilon(x/x_L)x + \epsilon u + x\eta,
\]
performing similar computations to those made before, we obtain the following binary constrained transition 
\[
x^\prime = x + \dfrac{\nu_\epsilon}{\epsilon^2 +\nu_\epsilon} \Phi^\epsilon(x/x_L)x - \dfrac{\epsilon^2}{\epsilon^2 + \nu_\epsilon}(x-x_d) + x\eta.
\]
Hence, we may proceed as explained in Section \eqref{grazing} to obtain in the regime $\epsilon\ll1$ and under the scaling \eqref{scal2} the Fokker-Planck equation  
\[
\partial_{t} \tilde f_a (x,t)= \partial_x\left\{ \left[   \dfrac{\mu}{2\delta} \left( \left( \dfrac{x}{x_L} \right)^\delta -1\right) x + \dfrac{x-x_d}{\nu}\right]\tilde f_{a}(x,t) + \dfrac{\sigma}{2} \partial_x (x^2\tilde f_a(x,t))\right\}
\]
whose equilibrium distribution, in the case $\delta<0$ is given by 
\[
\tilde f_{a,\infty}(x) = C(x_L,x_d,\sigma,\nu) \left( \dfrac{x_L}x \right)^{{\gamma}/{|\delta|} +2} \exp\left\{ - \dfrac{\gamma}{\delta^2} \left( \left( \dfrac{x_L}x \right)^{|\delta|} -1 \right) \right\} \exp\left\{ - \frac{x-x_d}{\sigma\nu}\right\}. 
\]
Therefore, we may observe that action of the control is not capable to modify the tails of the distribution.

\end{remark}

\subsection{Multiplicative control and equilibrium distribution}\label{multi}
The multiplicative   case  \fer{eq:inter_cons} can be treated likewise. The Lagrangian is now
 \[
\mathcal L(u,x^\prime) = J(x^\prime,u) + \alpha \left \langle x^\prime - x - u \Phi^\e(x/x_L)x -  x\eta_\e \right \rangle,
\]
with $\alpha$ the Lagrange multiplier. 
The optimality conditions in this case read
\[
\begin{cases}
\partial_u \mathcal L(x^\prime, u) =  \nu_\e u - \alpha \Phi^\e(x/x_L) x = 0\\
\partial_{x^\prime} \mathcal L(x^\prime, u) = \left \langle x^\prime - x_d \right \rangle + \alpha = 0. 
\end{cases}
\]
These conditions yield the optimal control 
\begin{equation}\label{eq:uop}
u^* =  - \dfrac{\Phi^\e(x/x_L)x}{\nu_\e + (\Phi^\e(x/x_L)x)^2}(x-x_d).
\end{equation}

Now, plugging \eqref{eq:uop} into \eqref{eq:inter_cons} we obtain the following optimal constrained microscopic interaction model
\begin{equation}
\label{eq:mult}
x^\prime_* = x - \dfrac{(\Phi^\e(x/x_L)x)^2}{\nu_\e + (\Phi^\e(x/x_L)x)^2}(x-x_d) + x\eta_\e. 
\end{equation}
Note that, at variance with the additive control case, in which the constrained interaction \fer{eq:add} is a balance between a growth term and a decrease term, the action of the control is such that only a decrease is possible, apart from random fluctuations. 
We may consider, as in Section \ref{addi} , the limit of grazing interactions, by choosing  $\nu_\e = \e \nu$, where $\nu >0$. Using \fer{eq:Phi} we obtain
 \[
 \dfrac{(\Phi^\e(x/x_L)x)^2}{\nu \e + (\Phi^\e(x/x_L)x)^2} \approx \frac{x^2}\nu \left[ \dfrac{\mu}{2\delta} \left( \left( \dfrac{x}{x_L} \right)^\delta -1\right)\right]^2
   \]
Hence, in the limit $\epsilon \rightarrow 0^+$ we obtain the Fokker-Planck equation for the controlled density $f_m(x,t)$ 
in presence of a multiplicative control
 \[
\partial_{t} f_m(x,t) = \partial_x \left\{ \frac{x^2}\nu \left[ \dfrac{\mu}{2\delta} \left( \left( \dfrac{x}{x_L} \right)^\delta -1\right)\right]^2(x-x_d) f_m(x,{t}) +\dfrac{\sigma}{2} \partial_x (x^2 f_m(x,{t})) \right\}.
 \]
whose equilibrium distribution, for $\delta \ne -1$ or $\delta \ne -1/2$, takes the form 
\[
f_{m,\infty}(x) =  C(x_L,x_d,\sigma,\nu)\; x^{-2} \;\exp\left\{-\frac{2}{\sigma\nu} \left(\frac{\mu}{2\delta}\right)^2  A_\delta(x)\right\}. 
\]
where 
\[
A_\delta(x) = x \left(  \left( \dfrac{x}{2\delta+2}-\dfrac{x_d}{2\delta+1}  \right) \left( \dfrac{x}{x_L} \right)^{2\delta} + \left( \dfrac{2x_d}{\delta+1} -\dfrac{2x}{\delta+2} \right) \left( \dfrac{x}{x_L} \right)^{\delta} +\dfrac{x}{2}-x_d \right),
\]
and $ C(x_L,x_d,\sigma,\nu)>0$ is a normalization constant. It is worth to observe that in the case $\delta = -1$ we obtain the following equilibrium distribution
\[
f_{m,\infty}(x) = C(x_L,x_d,\sigma,\nu)  x^{-2-\alpha} \exp\left\{ -\dfrac{\mu^2}{2\sigma\nu}  \left[ -(2x_L+x_d)x + \dfrac{x^2}{2}+\dfrac{x_L^2x_d}{x} \right] \right\},
\]
with $\alpha = \dfrac{\mu^2}{2\sigma\nu}(x_L^2 + 2x_L\;x_d)$, which can be rewritten as follows
\[
\begin{split}
f_{m,\infty}(x) &= C(x_L,x_d,\sigma,\nu) x^{-2-\alpha} \mathcal N\left(2x_L+x_d,\dfrac{2\sigma\nu}{\mu^2}\right) \chi(x\ge 0) \times \\
&\qquad\times \exp\left\{ -\dfrac{\mu^2}{2\sigma\nu} \left[ - \dfrac{(2x_L^2+x_d)^2}{2} + \dfrac{x_L^2x_d}{x}\right]\right\},
\end{split}\]
which exhibits therefore slim tails.

\begin{figure}
\centering
\includegraphics[scale = 0.35]{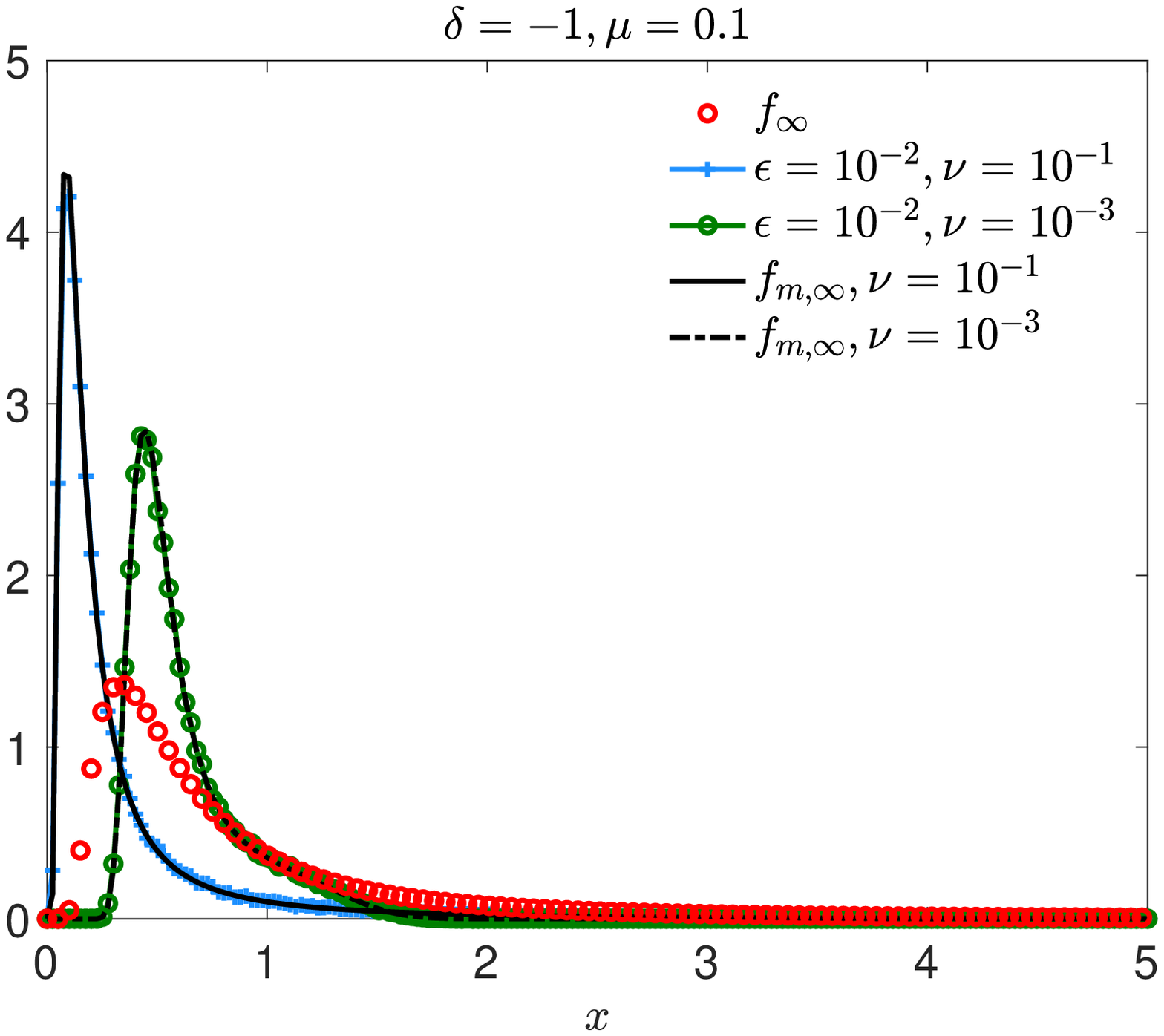}
\includegraphics[scale = 0.35]{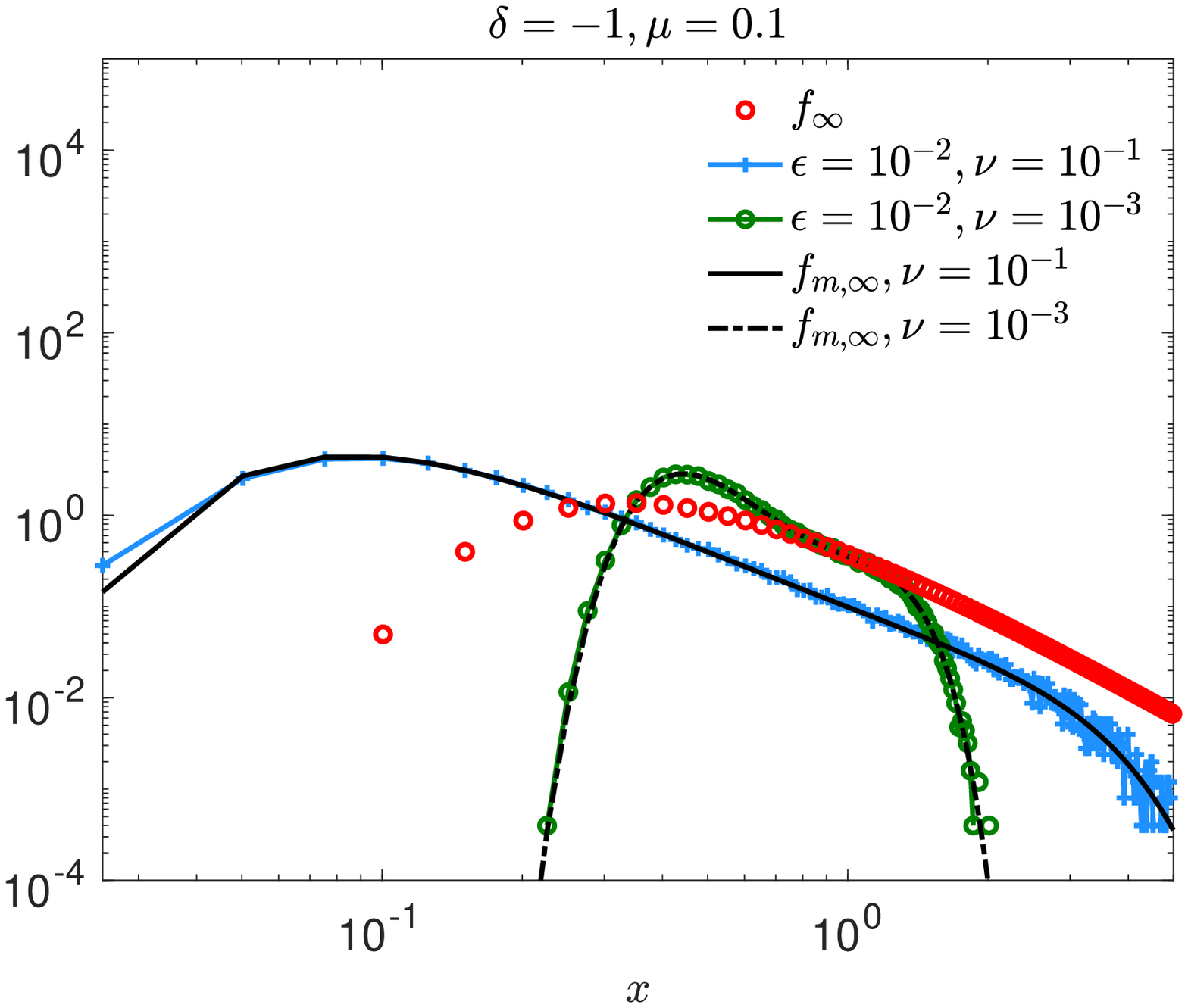}\\
\includegraphics[scale = 0.35]{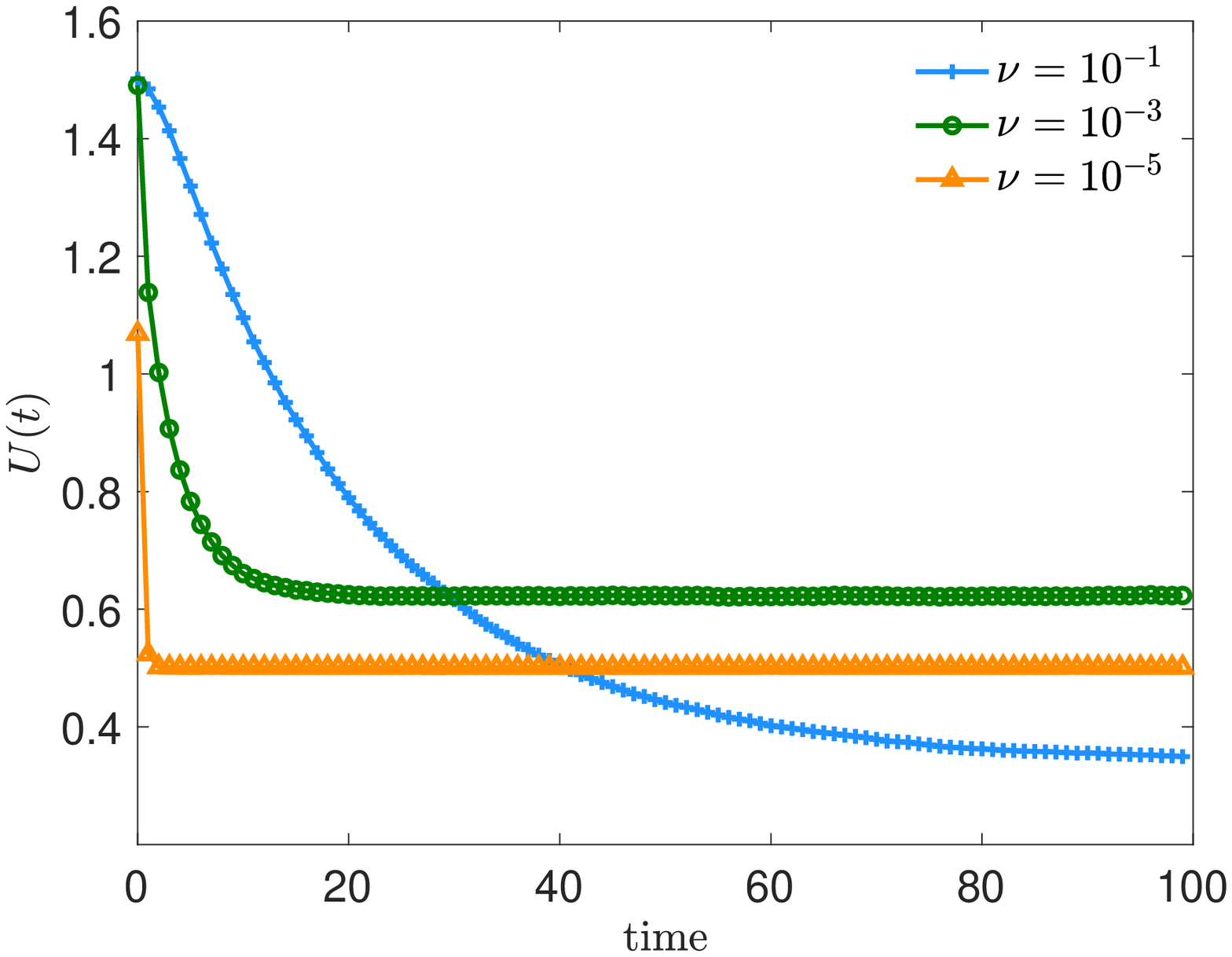}
\includegraphics[scale = 0.35]{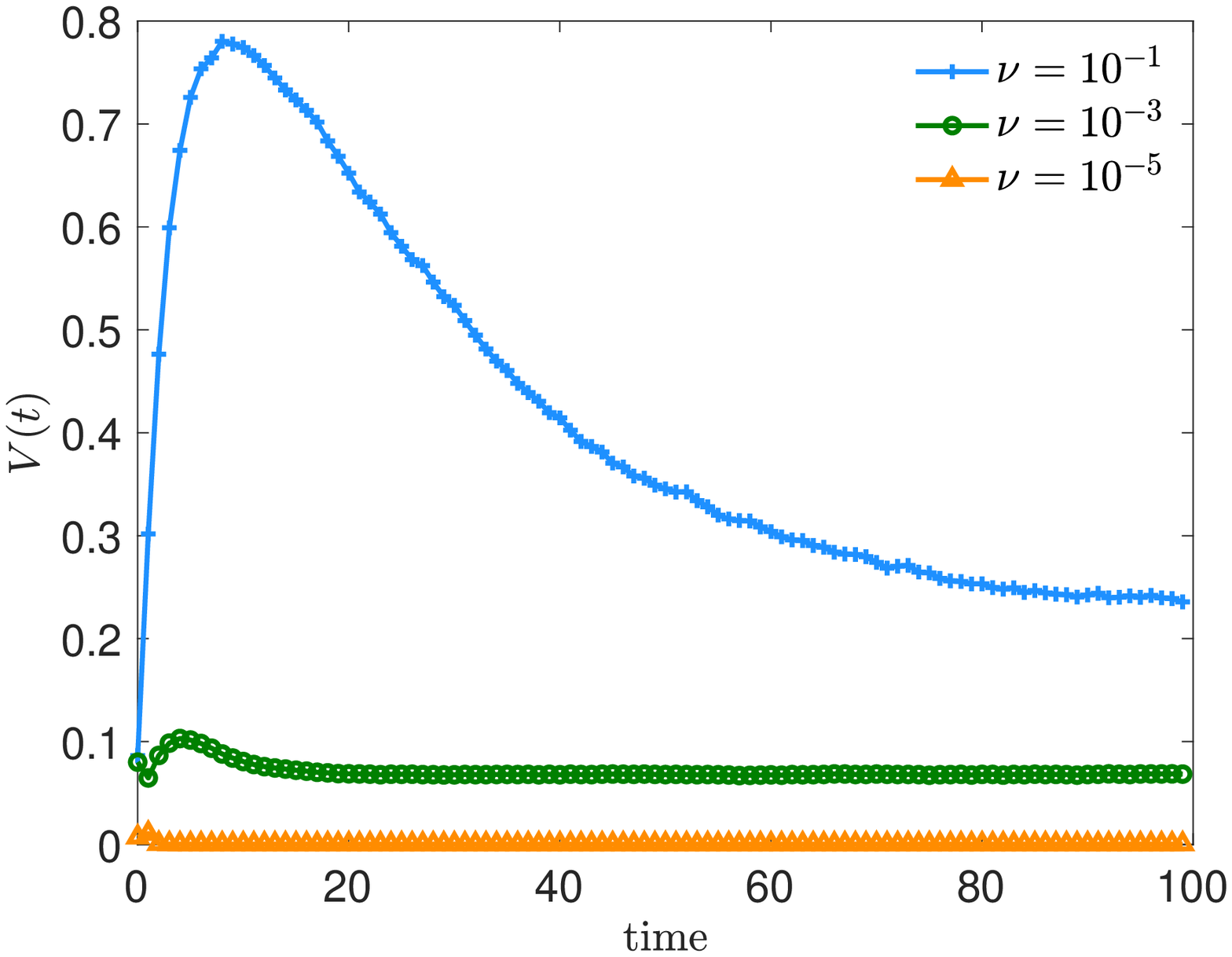}
\caption{Top row: comparison of the analytical steady states \eqref{prod} with the numerical solution of the Boltzmann-type equation for large times with multiplicative constrained interaction \eqref{eq:mult} in the quasi-invariant regime for $\epsilon = 10^{-2}$ in the cases $\nu=10^{-1},10^{-3}$.  Bottom row: evolution of $U(t)$, $V(t)$ defined in \eqref{eq:UV} for several values of the penalization coefficient. We considered $\lambda = \mu = 0.1$, $\sigma= 0.2$, $x_L = 1$ and target state $x_d =0.5$. In red dotted we report the equilibrium distribution of the unconstrained case. }
%{\bf Cosa succederebbe se si mettesse (per simmetria con la figura precedente) $\mu=0.9$? Perch\'e non s sono fatti media e varianza in quella figura? Ha senso mettere la curva senza controllo qui?}}
\label{fig:mult}
\end{figure}

In the top row of Figure \ref{fig:mult} we compare the numerical solution of the Boltzmann-type model \eqref{eq:boltz2} for large times with multiplicative control \eqref{eq:inter_cons} in the quasi-invariant regime $\epsilon = 10^{-2}$ and $\delta = -1$, $\mu = 0.1$. We considered several values of the penalization $\nu = 10^{-1},10^{-2}$ and a target size $x_d = \frac{1}{2}$ whereas $x_L = 1$. As before, the large time distribution of the Boltzmann model is consistently approximated by the ones of the Fokker-Planck regime for $\epsilon\ll1$. The action of the control is capable to modify the tails of the emerging distribution as highlighted by direct comparison with the unconstrained case. 

In order to better understand the effects of the multiplicative control we can look at the evolution of the mean size and of its variance, i.e. to the quantities
\begin{equation}
\label{eq:UV}
U(t) = \int_0^{+\infty} xf(x,t)dx, \qquad V(t) = \int_0^{+\infty} (x-U(t))^2f(x,t)dx.
\end{equation}
In the bottom row of Figure \ref{fig:mult} we report the evolution of $U(t)$ and $V(t)$ for several choices of $\nu>0$. We can observe how the control is capable to drive the expected size towards $x_d$ and to reduce the variance for small values of the penalization.

\section*{Conclusion}

In this paper we started by presenting a kinetic model for the distribution of tumor size and the related Fokker-Planck equation that yields under suitable ranges of parameters the most common growth laws used to characterize tumor growth.
We then showed that the emerging equilibrium distributions of the kinetic model {  correspond to} radically heterogeneous behaviors in terms of the decay of the tails,  according to the parameters of the model giving rise to the different growth laws. For instance, logistic-type growths are associated to a generalized Gamma density function characterized by slim tail with exponential decay. Gompertzian growth is associates to lognormal-type equilibria which exhibit slim tails as well. On the other hand, von Bertalanffy-type growths are associated to Amoroso-type distributions characterized by fat tails with polynomial decay. 

Now, from the pathological point of view fat-tailed distributions are related to a higher probability of finding large tumors with respect to thin-tailed distributions. So, from a therapeutical point of view it would be desirable to control at least the distribution tails. With this aim in mind we proved that optimal controls proportional to the tumor size acting either in an additive way or in a multiplicative way on the size transition function $\Phi^\e$ are able to do that. 

{  Extension of the proposed modelling approach to include the effects of the environment on the transition function and the possibility to consider more parsimonious controls are actually under study and will be discussed in a future work. }

\appendix

%For the sake of completeness, we will present in this appendix the principal tumor growth laws, which in a quantitative fashion describe the increase in tumor mass over time. These models are usually formulated in terms of differential equations that relate the growth rate of the tumor to its current state. Starting from these laws, and taking into account that growth can also be subject to random fluctuations, one can arrive to the same Fokker--Planck type equation \fer{eq:FP1}, thus showing the consistency of the present approach to others present in the literature, and the power of the kinetic approach, which allows to introduce the control at the level of the elementary interaction, a possibility which is not allowed in other approaches.

\section{Modelling tumor growth by ODEs}\label{app:A}

In the biomathematical literature  a variety of models for tumor growth have been proposed. The list  is quite long and the interested reader can have an almost complete picture about them by reading some exhaustive review papers  \cite{Gerlee,Norton,RCM,Ricciardi,West,Wheldon,Vaidya}. These essential growth models aim to catch the main features of the dynamics, often allowing to determine an analytical expression of  the evolution of the total number of cells in a tumor. 
%In order to obtain a statistical description of these dynamics, in the following we will present also a formal derivation of mean-field type equations. In this case the evolution of the distribution of tumors with a certain size is based on microscopic dynamics ruling the drift. 
In order to draw a comparison between the { classical and the present} approaches, in the following we will briefly recall the main features of a large class of { growth} models, for future reference. 

Most of the well-know models present in the literature can be described in a unified version by the class of first-order differential equations of Bernoulli type for the number $x(t)$ of tumor cells
 \be\label{gen}
 \dot x(t) = \frac\alpha\delta\, x(t) \left( 1 - \left(\frac{x(t)}{x_L}\right)^\delta \right),
 \ee
parameterized by  $\alpha >0$, $\delta\in[-1,1]$, and the  carrying capacity  $x_L$ of the system. If $\delta \not=0$, \eqref{gen} can be easily integrated to get the analytical solution
 \be\label{solu}
 x(t) = x_L \left\{  \left[  \left(\frac{x_L}{x_0} \right)^\delta  - 1  \right] e^{-\alpha t}  +1  \right\}^{-1/\delta}.
 \ee
describing the evolution of tumor cells starting from their initial number $x_0$ at time $t = 0$ toward the stable equilibrium represented by the carrying capacity $x_L$.
%Note that this solution holds independently of the fact that the initial value $x_0$ is less than $x_L$. However, if $x_0 > x_L$ we are in a situation in which the solution $x(t)$ is decreasing towards the limit value $x_L$. For this reason, growth models always require that $x_0 <x_L$. 

Equation \fer{gen} includes a variety of well-known growths like the \emph{logistic}, \emph{von Bertalanffy} and \emph{Gompertz growths}. Each of them has been widely used to describe the evolution tumours' size, see \cite{MVPF,West,Wheldon}.
%The logistic growth has been proposed as a correction of Malthus growth model, which assumes a pure exponential growth
% \[
% x(t) = x_0 e^{\alpha {t}},
% \]
% that does not take into account the finite capacity of a system. 
In fact, the logistic growth corresponds to fixing $\delta = 1$ in \eqref{gen} yielding
\be\label{logi}
\dot x(t) = \alpha x \left(1-\dfrac{x(t) }{x_L}  \right),
\ee
whose solution can be expressed in the form
\[
x(t) = \frac{x_L}{1+K x_L e^{-\alpha t}},
\]
 with $K = 1/x_0 - 1/x_L$. This growth model converges exponentially at the rate $\alpha$  towards the carrying capacity of the system $x_L$,  and it has been fruitfully employed in many applications in population dynamics. 
 Other logistic-type growth models correspond to the positive values of $\delta\in(0,1)$.

In the context of biological processes other models seem to furnish a better explanation about  real data of tumor growth \cite{HCMB}. These growth models belong to the class \fer{gen}, and are characterized by negative values of the constant $\delta$. The most known model in this range of the parameter is due to von Bertalanffy, and it is usually written in the form
\begin{equation}\label{eq:vonB}
\dot x(t) = p x(t)^a - qx(t), 
\end{equation}
where $0\le a<1$, and  $p,q>0$ are the rates of growth and size-proportional catabolism, respectively. This model corresponds to the choice $\delta = a-1 <0$, $\alpha = q(1-a)$ and  $x_L = \left({p}/{q} \right)^{{1}/{(1-a)}}$ in \fer{gen}. Substituting these values into \fer{solu}, its solution 
\[
x(t) = x_L \left[ 1-\left(1-\frac pq \,{x_0}^{1-a} \right) e^{-q(1-a)t } \right]^{\frac{1}{1-a}}, 
\]
converges exponentially fast at a rate $q$ towards  $x_L$. 

Finally,  the limit case $\delta \to 0$ in \fer{gen} corresponds to Gompertz growth. This growth is given as the solution of the differential equation
\begin{equation}\label{eq:gomp}
\dot x (t) = - \alpha x(t) \log \left( \dfrac{x(t)}{x_L} \right), 
\end{equation}
In \fer{eq:gomp} the constant  $\alpha>0$ denotes the growth rate related to the proliferative ability of cells. The exact solution of the Gompertz growth model can be easily found to be
\[
x(t) =  x_L \exp\left\{ e^{-\alpha t}  \log\dfrac{x_0}{x_L} \right\}.
\]
As in the previous cases  $\lim_{t\rightarrow +\infty} x(t) = x_L$  exponentially.

%\begin{figure}
%\centering
%\includegraphics[scale = 0.4]{figure/compare_micro_alpha02}
%%\includegraphics[scale = 0.4]{figure/compare_micro_alpha04}
%\caption{Evolution of the three presented growth models in the case $x_L = 1$, and $\alpha = 0.2$. We assumed $x_0 = 10^{-2}$, and for the von Bertalanffy model $q = p = 1$, and $a = 1-\alpha$.}
%\label{fig:evo_micro}
%\end{figure}
%
%In Figure \ref{fig:evo_micro} we compared the evolution presented models over the time interval $[0,50]$ for the regime $\alpha = 0.2$ and a fixed carrying capacity $x_L = 1$. We further assumed that for the von Bertalanffy model $p = q = 1$ and $1-a = \alpha$. 
 
 \section{Mean-field approaches for tumour growth containment}\label{app:B}
 
In the following we will present a formal derivation of mean-field type equations. In this case the evolution of the distribution of tumors with a certain size is based on microscopic dynamics ruling the drift. 
 
The deterministic dynamics of growth driven by equation \fer{gen} is the starting point to obtain partial differential equations able to describe the evolution of the density function $f(x,t)$ which, at a certain time $t=t_0$ measures the statistics of the size $x \ge 0$ of tumors which are growing according to \fer{gen} in a certain group of observed patients. Let $X(t)$,  denote the process which gives the statistical distribution of the sizes of tumors in the group at time  $t \ge 0$, and let $F(x,t)$ denote its distribution, defined by 
 \[
 F(x,t) = P(X(t) \le x), \qquad x \ge 0.
 \]
The classical way to recover the evolution of $F(x,t)$ consequent to a growth driven by equation \fer{gen} is to remark that, if $x(t)$ denotes the solution \fer{solu} to equation \fer{gen} departing from the value $x \ge 0$ at time $t_0 < t$, then 
 \[
 P(X(t) \le x(t)) = P(X(t_0) \le x),
 \]
or, what is the same
 \be\label{con2}
 F(x(t),t) = F(x,t_0) = const.
 \ee
Hence, taking the time derivative on both sides of \fer{con2} we obtain
 \be\label{con3}
 \left.\frac d{dt} F(x(t),t) = \frac{\partial F(x,t)}{\partial t} + \dot x(t)\frac{\partial F(x,t)}{\partial x}\right|_{x =x(t)} = 0.
 \ee
Using \fer{gen} into \fer{con3} one shows that $F(x,t)$  satisfies the conservation law \cite{Perth}
\be\label{con4}
 \frac{\partial F(x,t)}{\partial t} +\frac\alpha\delta\,x\, \left( 1 - \left(\frac{x}{x_L}\right)^\delta \right)\frac{\partial F(x,t)}{\partial x}= 0.
 \ee
Let us suppose that $F(x,t)$ is regular with respect to $x$, and let $f(x,t)$ denote the probability density of the process $X(t)$. In terms of the probability density $f(x,t)$ the conservation law is rewritten as
\be\label{con5}
 \frac{\partial f(x,t)}{\partial t} +\frac\alpha\delta\,\frac{\partial}{\partial x}\left[x\, \left( 1 - \left(\frac {x}{x_L}\right)^\delta \right)f(x,t) \right]= 0,
 \ee
which is obtained from \fer{con4} simply by differentiation with respect to $x$.  The complete description of the dynamics of growth is then obtained by taking into account that growth can also be subject to random fluctuations, which is reasonable to assume proportional to the size $X(t)$. This is classically obtained by introducing the multiplicative action on $X(t)$ of a standard Brownian motion of width $\sigma$, independent of $X(t)$ (cf. \cite{AG} and the references therein), which leads to adding  a second-order term into \fer{con5}. 
Thus, the resulting model is the Fokker--Planck type equation 
 \[
  \frac{\partial f(x,t)}{\partial t}= \frac {\sigma} 2 \frac{\partial^2 }{\partial x^2}
 \left(x^{2} f(x,t)\right )- \frac\alpha\delta\,\frac{\partial}{\partial x}\left[x\, \left( 1 - \left(\frac {x}{x_L}\right)^\delta \right)f(x,t) \right],
 \]
As an example, the Gompertz growth case $\delta \to 0$ considered in \cite{AG} is described by 
\be\label{FP-AG}
  \frac{\partial f(x,t)}{\partial t}= \frac {\sigma} 2 \frac{\partial^2 }{\partial x^2}
 \left(x^{2} f(x,t)\right )+ \alpha\,\frac{\partial}{\partial x}\left(x\log \frac{x}{x_L}\, f(x,t) \right).
 \ee
Once the growth model has been formalized, the effects of a given therapy is included in the model by assuming that the growth parameters in equation \fer{gen} are time-dependent functions \cite{AG}. In this way, the study of the growth in presence of a treatment can be approached by studying the modifications induced in time by these functions. Clearly, the knowledge of the action of these functions should allow to evaluate the effectiveness of the therapy on time, and in addition to better establish  treatment schedules. In the notations used in \cite{AG}, the parameters $\alpha$ and $x_L$ in the drift term of the Fokker--Planck equation \fer{FP-AG} have been considered as functions of time with the following dependence
 \be\label{coe}
 \alpha(t) = \bar \alpha - D(t), \quad x_L(t) = x_L \, \exp\left\{ - \frac{C(t)}{\bar\alpha - D(t)} \right\}.
 \ee 
where $C(t)$ and $D(t)$ (the therapy) have to be estimated to diminish at best the size of the tumor in time. Then, the strategy consists in performing experimental studies to test the effectiveness of the therapeutic treatment  including a control (untreated) group and one (or more) treated groups, where the growth in time of the control group follows the dynamics of the Fokker--Planck equation \fer{FP-AG}, while the treated groups are described by the modified Fokker--Planck equation \fer{FP-AG} in which the coefficients of the drift term are modified according to \fer{coe}. The comparison allows to estimate the unknown functions $C(t)$ and $D(t)$.

While this procedure helps to shed a light into the problem of finding the effects of the therapy, the choice of acting on growth in terms of the functions $C(t)$ and $D(t)$, which in the original formulation in \cite{AG} is additive, is largely arbitrary, and in any case does not help to find the best way to act on the growth to obtain regression, nor to understand the statistical variations on the resulting final distribution of the treated group with respect to the one of the untreated group.

\section*{Acknowledgement}
This work has been written within the activities of GNFM and GNCS groups of INdAM (National Institute of High Mathematics). The research of G. T. and M. Z. was partially supported by MIUR - Dipartimenti di Eccellenza Program (2018-2022) - Dept. of Mathematics  "F. Casorati" University of Pavia. The research of L.P. was partially supported by  MIUR - Dipartimenti di Eccellenza Program (2018-2022) - Dept. of Mathematical Sciences "G. L. Lagrange", Politecnico di Torino. 

%\vfill\eject

\end{document}